\documentclass[12pt,a4paper]{article}

\usepackage{ifthen} 
\usepackage{booktabs} 
\newboolean{pdflatex}
\setboolean{pdflatex}{true} 

\newboolean{articletitles}
\setboolean{articletitles}{true} 

\newboolean{uprightparticles}
\setboolean{uprightparticles}{false} 

\newboolean{inbibliography} 
\setboolean{inbibliography}{false} 

\usepackage{microtype}
\usepackage{lineno}  % for line numbering during review
\usepackage{xspace} % To avoid problems with missing or double spaces after
\usepackage{caption} %these three command get the figure and table captions automatically small

\usepackage{graphicx}  % to include figures (can also use other packages)
\usepackage{color}
\usepackage{colortbl}
\graphicspath{{./figs/}} % Make Latex search fig subdir for figures

\usepackage{amsmath} % Adds a large collection of math symbols
\usepackage{amssymb}
\usepackage{amsfonts}
\usepackage{upgreek} % Adds in support for greek letters in roman typeset

\newcommand*\patchAmsMathEnvironmentForLineno[1]{%
\expandafter\let\csname old#1\expandafter\endcsname\csname #1\endcsname
\expandafter\let\csname oldend#1\expandafter\endcsname\csname
end#1\endcsname
 \renewenvironment{#1}%
   {\linenomath\csname old#1\endcsname}%
   {\csname oldend#1\endcsname\endlinenomath}%
}
\newcommand*\patchBothAmsMathEnvironmentsForLineno[1]{%
  \patchAmsMathEnvironmentForLineno{#1}%
  \patchAmsMathEnvironmentForLineno{#1*}%
}
\AtBeginDocument{%
\patchBothAmsMathEnvironmentsForLineno{equation}%
\patchBothAmsMathEnvironmentsForLineno{align}%
\patchBothAmsMathEnvironmentsForLineno{flalign}%
\patchBothAmsMathEnvironmentsForLineno{alignat}%
\patchBothAmsMathEnvironmentsForLineno{gather}%
\patchBothAmsMathEnvironmentsForLineno{multline}%
\patchBothAmsMathEnvironmentsForLineno{eqnarray}%
}

\usepackage{hyperref}    % Hyperlinks in references
\usepackage[all]{hypcap} % Internal hyperlinks to floats.

\def\pt         {\mbox{$p_{\rm T}$}\xspace}
\newcommand{\tev}{\ifthenelse{\boolean{inbibliography}}{\ensuremath{~T\kern -0.05em eV}\xspace}{\ensuremath{\mathrm{\,Te\kern -0.1em V}}}\xspace}
\newcommand{\gev}{\ensuremath{\mathrm{\,Ge\kern -0.1em V}}\xspace}
\newcommand{\stat}{\ensuremath{\mathrm{\,(stat)}}\xspace}
\newcommand{\syst}{\ensuremath{\mathrm{\,(syst)}}\xspace}
\def\fb   {\ensuremath{\mbox{\,fb}}\xspace}

\def\lhcb {\mbox{LHCb}\xspace}
 \def\Pb      {\ensuremath{\mathrm{b}}\xspace}                 
 \def\Pc      {\ensuremath{\mathrm{c}}\xspace} 
\def\bquark    {{\ensuremath{\Pb}}\xspace}
\def\cquark    {{\ensuremath{\Pc}}\xspace}
\def\pythia     {\mbox{\textsc{Pythia}}\xspace}
\def\evtgen     {\mbox{\textsc{EvtGen}}\xspace}
\def\photos     {\mbox{\textsc{Photos}}\xspace}
\def\geant      {\mbox{\textsc{Geant4}}\xspace}
\def\ptot       {\mbox{$p$}\xspace}
\def\mum  {\ensuremath{{\,\upmu\rm m}}\xspace}
\def\mm   {\ensuremath{\rm \,mm}\xspace}

\usepackage{cite} % Allows for ranges in citations
\usepackage{mciteplus}
\usepackage{longtable}
\usepackage{chngpage}

\def \light {\ensuremath{\rm light\mbox{-}parton}\xspace}

\def \bdtbcl {\ensuremath{{\rm BDT}(bc|udsg)}\xspace}
\def \bdtbc {\ensuremath{{\rm BDT}(b|c)}\xspace}
\def \muptr {\ensuremath{\pt(\mu)/\pt(j_\mu)}\xspace}
\def \wc {\ensuremath{W\!+\!c}\xspace}
\def \wb {\ensuremath{W\!+\!b}\xspace}

\begin{document}

\renewcommand{\thefootnote}{\fnsymbol{footnote}}
\setcounter{footnote}{1}

\begin{titlepage}
\pagenumbering{roman}

% Header ---------------------------------------------------
\vspace*{-1.5cm}
\centerline{\large EUROPEAN ORGANIZATION FOR NUCLEAR RESEARCH (CERN)}
\vspace*{1.5cm}
\hspace*{-0.5cm}
\begin{tabular*}{\linewidth}{lc@{\extracolsep{\fill}}r}
\ifthenelse{\boolean{pdflatex}}
{\vspace*{-2.7cm}\mbox{\!\!\!\includegraphics[width=.14\textwidth]{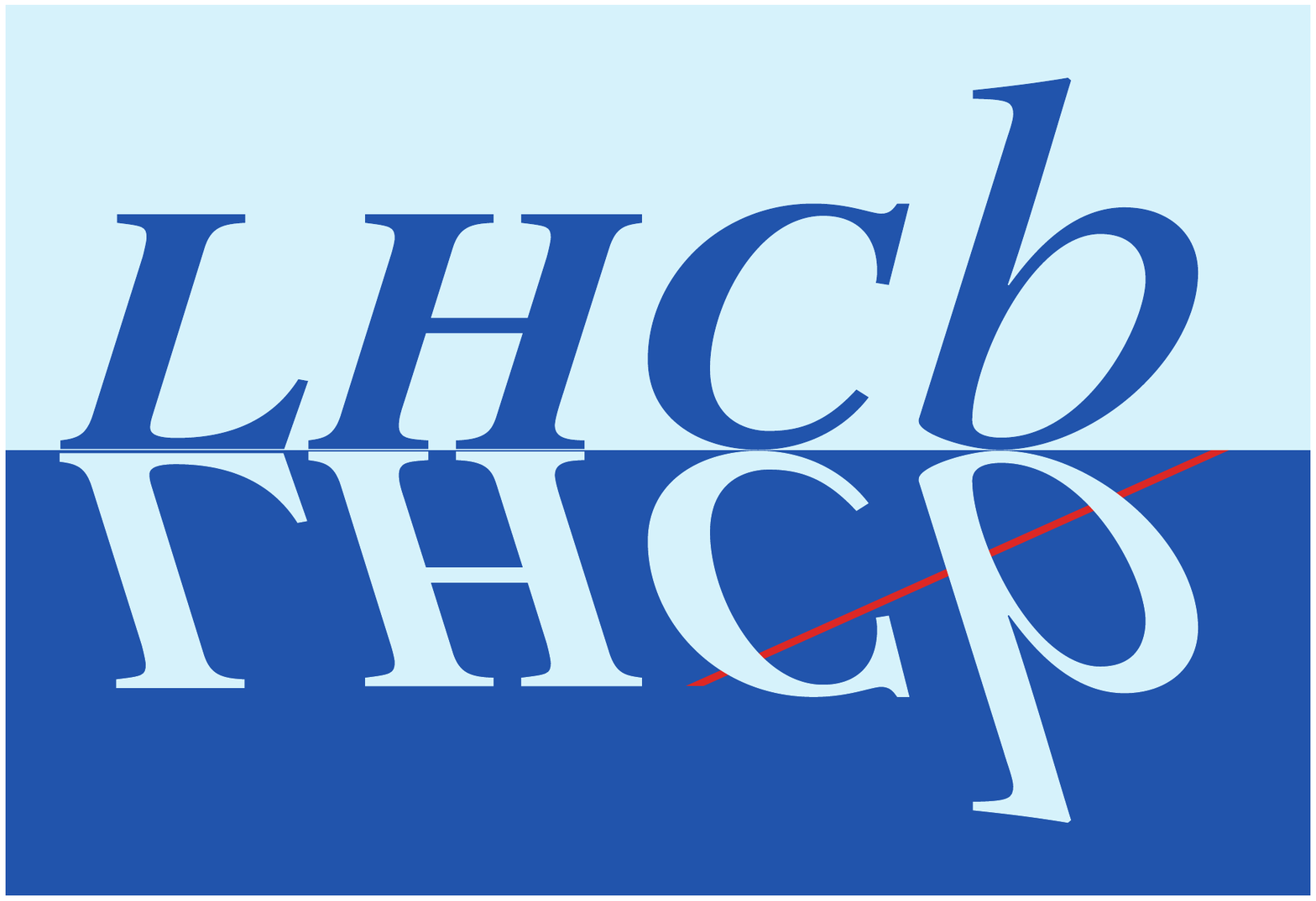}} & &}%
{\vspace*{-1.2cm}\mbox{\!\!\!\includegraphics[width=.12\textwidth]{lhcb-logo.eps}} & &}%
\\
 & & CERN-PH-EP-2015-118 \\ 
 & & LHCb-PAPER-2015-021 \\  
 & & September 8, 2015\\
 & & \\
\end{tabular*}

\vspace*{4.0cm}

% Title --------------------------------------------------
{\bf\boldmath\huge
\begin{center}
  Study of $W$ boson production in association with beauty and charm 
\end{center}
}

\vspace*{0.5cm}

% Authors -------------------------------------------------
\begin{center}
The LHCb collaboration\footnote{Authors are listed at the end of this article.}
\end{center}

\vspace{\fill}

% Abstract -----------------------------------------------
\begin{abstract}
  \noindent
The associated production of a $W$ boson with a jet originating from either a light parton or heavy-flavor quark is studied  in the forward region using proton-proton collisions. The analysis uses data corresponding to integrated luminosities of 1.0 and $2.0\fb^{-1}$ collected with the LHCb detector at center-of-mass energies of 7 and 8\tev, respectively. The $W$ bosons are reconstructed using the $W\to\mu\nu$ decay and muons with a transverse momentum, \pt, larger than 20\gev in the pseudorapidity range $2.0<\eta<4.5$. The partons are reconstructed as jets with $\pt > 20\gev$ and $2.2 < \eta < 4.2$. The sum of the muon and jet momenta must satisfy $\pt > 20\gev$. The fraction of $W+$jet events that originate from beauty and charm quarks is measured, along with the charge asymmetries of the $W\!+\!b$ and $W\!+\!c$ production cross sections. The ratio of the $W+$jet to $Z+$jet production cross sections is also measured using the $Z\to\mu\mu$ decay. All results are in agreement with Standard Model predictions.
\end{abstract}

\vspace*{0.5cm}

\begin{center}
  Phys.\ Rev.\ D.\ {\bf 92} (2015) 052001
\end{center}

\vspace{\fill}

{\footnotesize 
\centerline{\copyright~CERN on behalf of the \lhcb collaboration, license \href{http://creativecommons.org/licenses/by/4.0/}{CC-BY-4.0}.}}
\vspace*{2mm}

\end{titlepage}

\newpage
\setcounter{page}{2}
\mbox{~}
\newpage

\renewcommand{\thefootnote}{\arabic{footnote}}
\setcounter{footnote}{0}

\pagestyle{plain} % restore page numbers for the main text
\setcounter{page}{1}
\pagenumbering{arabic}

%\linenumbers

\clearpage

\section{Introduction}

Measurements of $W+$jet production in hadron collisions provide important tests of the Standard Model (SM), especially of perturbative quantum chromodynamics (QCD) in the presence of heavy-flavor quarks.  Such measurements are also sensitive probes of the parton distribution functions (PDFs) of the proton. The ratio of the $W+$jet to $Z+$jet production cross sections is a test of perturbative QCD methods and constrains the light-parton PDFs of the proton.

The jet produced in association with the $W$ boson may originate either from a $b$ quark (\wb), $c$ quark (\wc) or light parton.
Several processes contribute to the \wb and \wc final states at next-to-leading order (NLO) in perturbative QCD. The dominant mechanism for \wc production is $gs \to Wc$, but there are also important contributions from $gs \to Wcg$, $gg \to Wc\bar{s}$, and $q\bar{q}\to Wc\bar{c}$\cite{Stirling:2012vh}. Therefore, measuring the ratio of the \wc to $W+$jet production cross sections  in the forward region at the LHC provides important constraints on the $s$ quark PDF~\cite{Baur:1993zd,Giele:1995kr} at momentum transfers of $Q \approx 100\gev$ ($c=1$ throughout this article) and momentum fractions down to $x \approx 10^{-5}$. Previous measurements of the proton $s$ quark PDF were primarily based on deep inelastic scattering experiments with $Q \approx 1\gev$ and $x$ values $\mathcal{O}(0.1)$~\cite{Mason:2007zz,Goncharov:2001qe,Airapetian:2013zaw}. 
The \wc cross section has been measured at the Tevatron~\cite{Abazov:2008qz,Aaltonen:2012wn} and at the LHC~\cite{Chatrchyan:2013uja,Aad:2014xca} in the central region.

In the so-called four-flavor scheme, theoretical calculations are performed considering only the four lightest quarks in the proton~\cite{Mangano:1992kp}.   Production of \wb proceeds via $q\bar{q} \to Wg$ with $g\to b\bar{b}$ at leading order. If the $b$ quark content of the proton is considered, {\it i.e.} the five-flavor scheme, then single-$b$ production via $qb \to W bq$ also contributes~\cite{mcfm.wb}. The ratio of the \wb to $W+$jet cross sections thus places constraints both on the intrinsic $b$ quark content of the proton and the probability of gluon splitting into $b\bar{b}$ pairs. 
The \wb cross section has been measured in the central region at the Tevatron~\cite{Aaltonen:2009qi,D0:2012qt} and at the LHC~\cite{Aad:2013vka}.

LHCb has measured the cross sections for inclusive $W$ and $Z$ production in proton-proton ($pp$) collisions at center-of-mass energy $\sqrt{s}=7\tev$~\cite{LHCb-PAPER-2014-033, LHCb-PAPER-2015-001, LHCb-PAPER-2012-036, LHCb-PAPER-2015-003}, providing precision tests of the SM in the forward region. Additionally, measurements of the $Z+$jet and $Z\!+\!b$ cross sections have been made\cite{LHCb-PAPER-2013-058,LHCb-PAPER-2014-055}. 
In this article, the associated production of a $W$ boson with a jet originating from either a light parton or a heavy-flavor quark is studied using $pp$ collisions at center-of-mass energies of 7 and 8\tev.
The production of the \wb final state via top quark decay is not included in the signal definition in this analysis, but is reported separately in Ref.~\cite{TOP}.

A comprehensive approach is taken, where the inclusive $W+$jet, \wb and \wc contributions are measured simultaneously, rather than split across multiple measurements as in Refs.~\cite{Khachatryan:2014uva,Chatrchyan:2013uja,Chatrchyan:2013uza,Aad:2014qxa,Aad:2014xca,Aad:2013vka,Aad:2014rta}. The identification of $c$ jets, in conjunction with $b$ jets, is performed using the tagging algorithm described in Ref.\cite{LHCb-PAPER-2015-016}, which improves upon previous $c$-tagging methods where muons or exclusive decays were required to identify the jet\cite{Chatrchyan:2013uja,Aad:2014xca}.
For each center-of-mass energy, the following production cross section ratios are measured: $\sigma(Wb)/\sigma(Wj)$, $\sigma(Wc)/\sigma(Wj)$, $\sigma(W^+j)/\sigma(Zj)$, $\sigma(W^-j)/\sigma(Zj)$, $\mathcal{A}(Wb)$, and $\mathcal{A}(Wc)$, where 
\begin{equation}
\mathcal{A}(Wq)\equiv \frac{\sigma(W^+q)-\sigma(W^-q)}{\sigma(W^+q)+\sigma(W^-q)}.
\end{equation}
The analysis is performed using the $W \to \mu\nu$ decay  and
jets clustered with the anti-$k_{\rm T}$ algorithm~\cite{1126-6708-2008-04-063} using a distance parameter $R=0.5$.
The following fiducial requirements are applied: both the muon and the jet must have momentum transverse to the beam, \pt, greater than 20\gev; the pseudorapidity of the muon must fall within $2.0<\eta(\mu)<4.5$; the jet pseudorapidity must satisfy $2.2<\eta(j)<4.2$; the muon and jet must be separated by $\Delta R(\mu,j) > 0.5$, where $\Delta R \equiv \sqrt{\Delta\eta^2 + \Delta\phi^2}$ and $\Delta\eta(\Delta\phi)$ is the difference in pseudorapidity (azimuthal angle) between the muon and jet momenta; and the transverse component of the sum of the muon and jet momenta must satisfy $\pt(\mu+j)  \equiv \left(\vec{p}(\mu) + \vec{p}(j)\right)_{\rm T} > 20\gev$.
All results reported in this article are for within this fiducial region, {\em i.e.}\ no extrapolation outside of this region is performed.

The article is organized as follows: the detector, data sample and simulation are described in Sect.~\ref{sec:data}; the event selection is given in Sect.~\ref{sec:selection}; the signal yields are determined in Sect.~\ref{sec:yields}; the systematic uncertainties are outlined in Sect.~\ref{sec:uncertainty}; and the results are presented in Sect.~\ref{sec:results}.

\section{The LHCb detector and data set}\label{sec:data}

The \lhcb detector~\cite{Alves:2008zz,Aaij:2014jba} is a single-arm forward
spectrometer covering the \mbox{pseudorapidity} range $2<\eta <5$,
designed for the study of particles containing \bquark or \cquark
quarks. The detector includes a high-precision tracking system
consisting of a silicon-strip vertex detector surrounding the $pp$
interaction region~\cite{LHCbVELOGroup:2014uea}, a large-area silicon-strip detector located
upstream of a dipole magnet with a bending power of about
$4{\rm\,Tm}$, and three stations of silicon-strip detectors and straw
drift tubes~\cite{LHCb-DP-2013-003} placed downstream of the magnet.
The tracking system provides a measurement of momentum, \ptot, of charged particles with
a relative uncertainty that varies from 0.5\% at low momentum to 1.0\% at 200\gev.
The minimum distance of a track to a primary vertex, the impact parameter, is measured with a resolution of $(15+29/\pt)\mum$,
with \pt in\,\gev.
Different types of charged hadrons are distinguished using information
from two ring-imaging Cherenkov detectors. 
Photons, electrons and hadrons are identified by a calorimeter system consisting of
scintillating-pad and preshower detectors, an electromagnetic
calorimeter and a hadronic calorimeter.
The electromagnetic and hadronic calorimeters have energy resolutions of $\sigma(E)/E = 10\%/\sqrt{E}\oplus 1\%$ and $\sigma(E)/E = 69\%/\sqrt{E}\oplus 9\%$ (with $E$ in GeV), respectively.
Muons are identified by a
system composed of alternating layers of iron and multiwire
proportional chambers~\cite{LHCb-DP-2012-002}.

The trigger~\cite{LHCb-DP-2012-004} consists of a
hardware stage, based on information from the calorimeter and muon
systems, followed by a software stage, which applies a full event
reconstruction. This analysis requires at least one muon candidate that satisfies the trigger requirement of $\pt>10\gev$. Global event cuts (GECs), which prevent high-occupancy events from dominating the processing time of the software trigger, are also applied and have an efficiency of about $90\%$ for $W+$jet and $Z+$jet events.

Two sets of $pp$ collision data collected with the LHCb detector are used: data collected during 2011 at $\sqrt{s}=7\tev$, corresponding to an integrated luminosity of 1.0~${\rm fb}^{-1}$, and data collected during 2012 at $\sqrt{s}=8\tev$, corresponding to an integrated luminosity of 2.0~${\rm fb}^{-1}$. Simulated $pp$ collisions, used to study the detector response, to define the event selection and to validate data-driven techniques, are generated using \pythia~\cite{Sjostrand:2006za,*Sjostrand:2007gs} with an \lhcb
configuration~\cite{LHCb-PROC-2010-056}. Decays of hadronic particles
are described by \evtgen~\cite{Lange:2001uf} in which final-state
radiation (FSR) is generated using \photos~\cite{Golonka:2005pn}. The
interaction of the generated particles with the detector, and its
response, are implemented using the \geant
toolkit~\cite{Allison:2006ve, *Agostinelli:2002hh} as described in
Ref.~\cite{LHCb-PROC-2011-006}.

Results are compared with theoretical calculations at NLO using MCFM~\cite{Campbell:2000bg} and the CT10 PDF set~\cite{Lai:2010vv}. The theoretical uncertainty is a combination of PDF, scale, and strong-coupling ($\alpha_s$) uncertainties. 
The PDF and scale uncertainties are evaluated following Refs.~\cite{Lai:2010vv} and \cite{Hamilton:2013fea}, respectively.
The $\alpha_s$ uncertainty is evaluated as the envelope obtained using $\alpha_s(M_Z) \in [0.117, 0.118, 0.119]$ in the theory calculations.

\section{Event selection}\label{sec:selection}

The signature for $W+$jet events is an isolated high-\pt muon and a well-separated jet, both produced in the same $pp$ interaction. 
Muon candidates are identified with tracks that have associated hits in the muon system.
The muon candidate must have $\pt(\mu) > 20\gev$ and pseudorapidity within ${2.0<\eta(\mu)<4.5}$. 
 Background muons from $W\to\tau\to\mu$ decays or semileptonic decays of heavy-flavor hadrons are suppressed by requiring the muon impact parameter to be less than $0.04\mm$~\cite{LHCb-PAPER-2014-033}. Background from high-momentum kaons and pions that enter the muon system and are misidentified as muons, is reduced by requiring that the sum of the energy of the associated electromagnetic and hadronic calorimeter deposits does not exceed 4\% of the momentum of the muon candidate.

Jets are clustered using the anti-$k_T$ algorithm  with a distance parameter $R=0.5$, as implemented in
\textsc{Fastjet}~\cite{fastjet}. 
Information from all the detector
subsystems is used to create charged and neutral particle inputs to
the jet-clustering algorithm using a particle flow approach~\cite{LHCb-PAPER-2013-058}. 
During 2011 and 2012, LHCb collected data with a mean number of $pp$ collisions per beam crossing of about 1.7.  To reduce contamination from multiple $pp$ interactions, charged particles reconstructed within the vertex detector may only be clustered into a jet if they are associated with the same $pp$ collision.

Signal events are selected by requiring a muon candidate and at least one jet with $\Delta R(\mu,j) > 0.5$. 
For each event the highest-\pt muon candidate that satisfies the trigger requirements is selected, along with the highest-\pt jet from the same $pp$ collision.  
The high-\pt muon candidate is not removed from the anti-$k_T$ inputs and so is clustered into a jet.
 This jet, referred to as the muon jet and denoted as $j_{\mu}$,
is used to discriminate between $W+$jet and dijet events.
The requirement $\pt(j_\mu+j) > 20\gev$ is made to suppress dijet backgrounds, which are well balanced in \pt, unlike $W+$jet events where there is undetected energy from the neutrino.  
Furthermore, the distribution of the fractional muon candidate \pt within the muon jet, $\pt(\mu)/\pt(j_{\mu})$, is used to separate vector bosons from jets.
For vector-boson production, this ratio deviates from unity only due to muon FSR, activity from the underlying event, or from neutral-particle production in a separate $pp$ collision, whereas for jet production this ratio is driven to smaller values by the presence of additional radiation produced in association with the muon candidate.

Events with a second, oppositely charged, muon candidate from the same $pp$ collision are vetoed. However, when the dimuon invariant mass is in the range $60 < M(\mu^+\mu^-) < 120\gev$, such events are selected as $Z+$jet candidates and the $\pt(j_{\mu}+j)$ requirement is not applied.
Two $Z+$jet data samples are selected at each center-of-mass energy: a data sample where only the $\mu^+$ is required to satisfy the trigger requirements and one where only the $\mu^-$ is required to satisfy them.  
The first sample is used to measure $\sigma(W^+j)/\sigma(Zj)$, while the second is used for $\sigma(W^-j)/\sigma(Zj)$.  This strategy leads to approximate cancellation of the uncertainty in the trigger efficiency in the measurement of these ratios.

The reconstructed jets must have $\pt(j) > 20\gev$ and $2.2 < \eta(j) < 4.2$. The reduced $\eta(j)$ acceptance ensures nearly uniform jet reconstruction and heavy-flavor tagging efficiencies.
The momentum of a reconstructed jet is scaled to obtain an unbiased
estimate of the true jet momentum. The scaling factor, typically between
0.9 and 1.1, is determined from simulation and depends on the jet \pt and $\eta$, the fraction of the jet transverse momentum measured with the
tracking systems, and the number of $pp$ interactions in the event.
No scaling is applied to the momentum of the muon jet.
Migration of events in and out of the jet \pt fiducial region due to the detector response is corrected for by an unfolding technique.  Data-driven methods are used to obtain the unfolding matrix, with the resulting corrections to the measurements presented in this article being at the percent level. 

The jets are identified, or tagged, as originating from the hadronization of a heavy-flavor quark by the presence of a secondary vertex (SV) with $\Delta R < 0.5$ between the jet axis and the SV direction of flight, defined by the vector from the $pp$ interaction point to the SV position. 
Two boosted decision trees (BDTs)~\cite{Breiman,Adaboost}, \bdtbcl and \bdtbc, trained on the characteristics of the SV and the jet, are used to 
 separate heavy-flavor jets from \light jets, 
and to separate $b$ jets from $c$ jets.  
The two-dimensional distribution of the BDT response observed in data is fitted to obtain the SV-tagged $b$, $c$ and \light jet yields. The SV-tagger algorithm is detailed in Ref.~\cite{LHCb-PAPER-2015-016}, where the heavy-flavor tagging efficiencies and \light mistag probabilities are measured in data.    

\section{Background determination}\label{sec:yields}

Contributions from six processes are considered in the $W+$jet data sample: $W+$jet signal events; $Z+$jet events where one muon is not reconstructed; top quark events producing a $W+$jet final state; $Z\to\tau\tau$ events where one $\tau$ lepton decays to a muon and the other decays hadronically; QCD dijet events; and vector boson pair production.  Simulations based on NLO predictions show that the last contribution is negligible.  

The signal yields are obtained for each muon charge and center-of-mass energy independently.  
The \muptr distribution is fitted to determine the $W+$jet yield of each data sample.  To determine the \wb and \wc yields, the subset of candidates with an SV-tagged jet is binned according to \muptr.  In each \muptr bin, the two-dimensional SV-tagger BDT-response distributions are fitted to determine the yields of $b$-tagged and $c$-tagged jets, which are used to form the \muptr distributions for candidates with $b$-tagged and $c$-tagged jets.  These \muptr distributions are fitted to determine the SV-tagged \wb and \wc yields.  Finally, to obtain $\sigma(Wb)/\sigma(Wj)$ and $\sigma(Wc)/\sigma(Wj)$, the jet-tagging efficiencies of $\epsilon_{\rm tag}(b) \approx 65\%$ and $\epsilon_{\rm tag}(c) \approx 25\%$ are accounted for.  In all fits performed in this analysis, the templates are histograms  with fixed shapes.

\begin{figure}
  \begin{center}
    \includegraphics[width=0.95\textwidth]{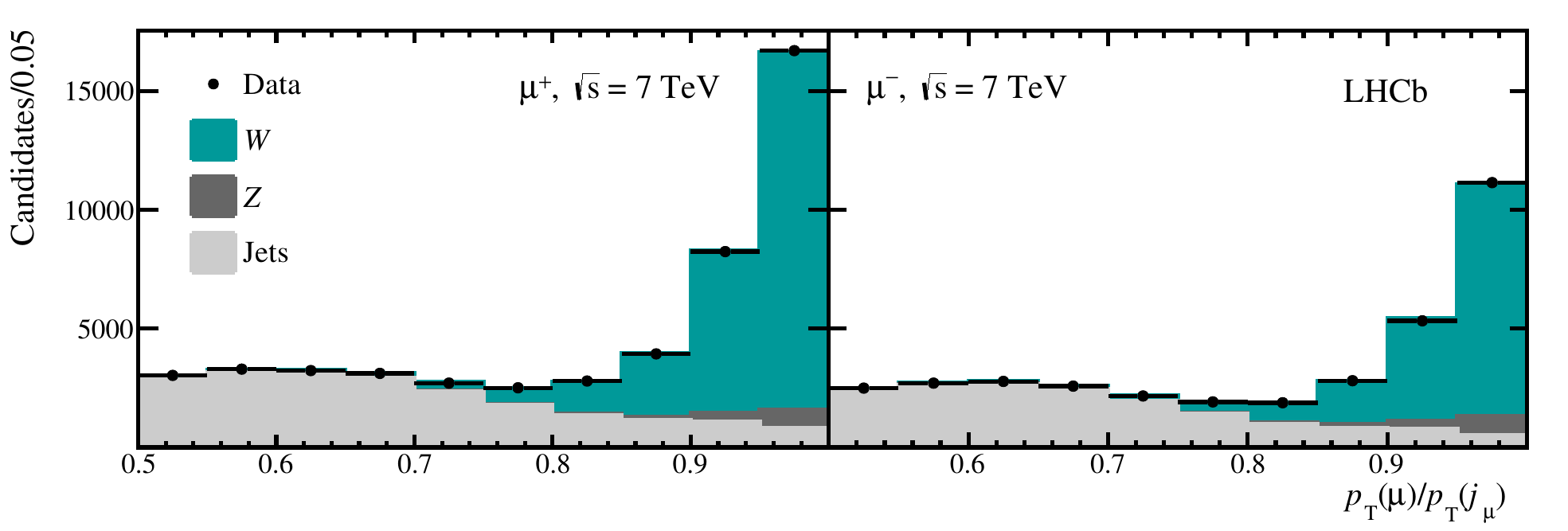}\\
    \vspace{-0.33in}
    \includegraphics[width=0.95\textwidth]{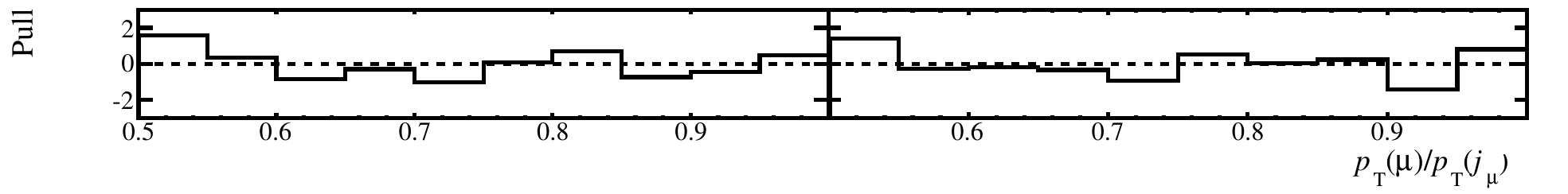}\\
    \includegraphics[width=0.95\textwidth]{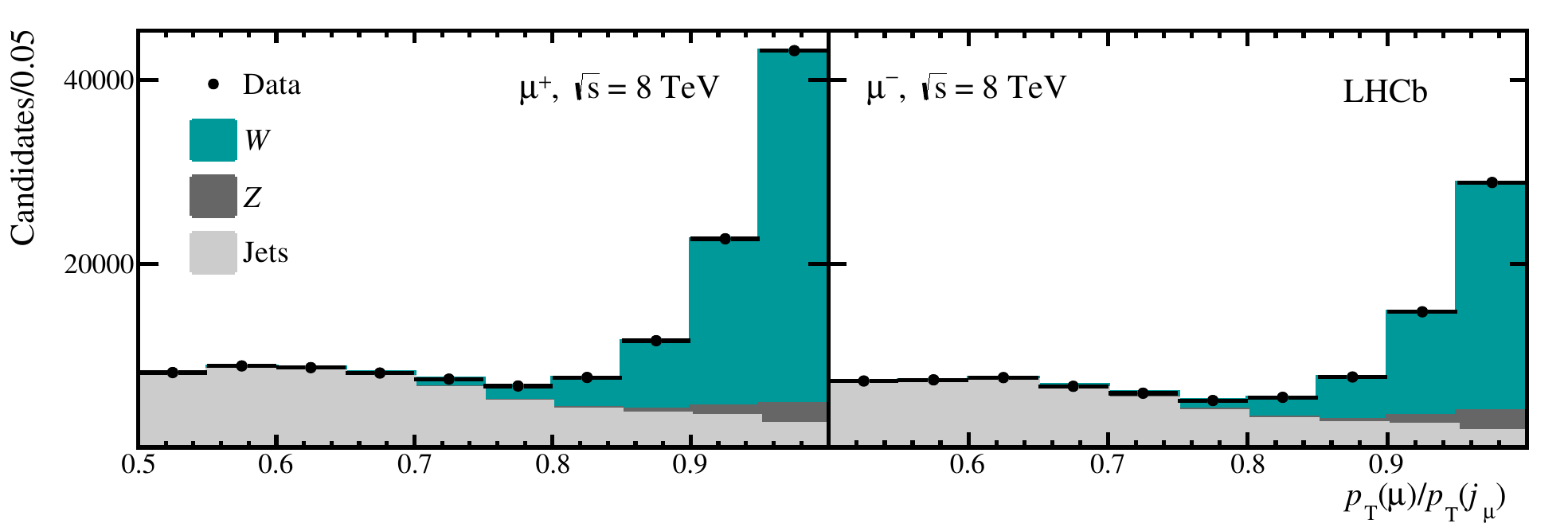}\\
    \vspace{-0.33in}
    \includegraphics[width=0.95\textwidth]{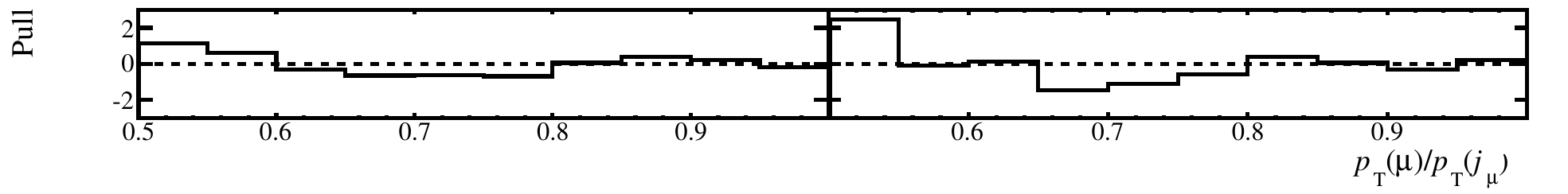}
    \caption{\label{fig:wmuptr_fit}  Distributions of \muptr with fits overlaid from (top) $\sqrt{s}=7\protect\tev$ and (bottom) $8\protect\tev$ data for (left) $\mu^+$ and (right) $\mu^-$.}
  \end{center}
\end{figure}

The \muptr distributions are shown in Fig.~\ref{fig:wmuptr_fit} (in this and subsequent figures the pull represents the difference between the data and the fit, in units of standard deviations). 
The $W$ boson yields are determined by performing binned extended-maximum-likelihood fits to these distributions with the following components:
\begin{itemize}
\item The $W$ boson template is obtained by correcting the \muptr distribution observed in $Z+$jet events for small differences between $W$ and $Z$ decays derived from simulation.
\item The template for $Z$ boson events where one muon is not reconstructed is obtained by correcting, using simulation, the \muptr distribution observed in fully reconstructed $Z+$jet events for small differences expected in partially reconstructed $Z+$jet events.
The yield is fixed from the fully reconstructed $Z+$jet data sample, where simulation is used to obtain the probability that the muon is missed, either because it is out of acceptance or it is not reconstructed.  
\item The templates for $b$, $c$ and \light jets are obtained using dijet-enriched data samples.  These samples require $\pt(j_{\mu}+j) < 10\gev$ and, for the heavy-flavor samples, either a stringent $b$-tag or $c$-tag requirement on the associated jet.  The templates are corrected for differences in the $\pt(j_{\mu})$ spectra between the dijet-enriched and signal regions.  The contributions of $b$, $c$ and \light jets are each free to vary in the \muptr fits.    
\end{itemize}
The \muptr fits determine the $W+$jet yields, which include contributions from top quark and $Z\to\tau\tau$ production. The top quark and $Z\to\tau\tau$ contributions cannot be separated from $W+$jet since their \muptr distributions are nearly identical to that of $W+$jet events. The subtraction of these backgrounds is described below.

\begin{figure}[]
  \begin{center}
\includegraphics[width=0.49\textwidth]{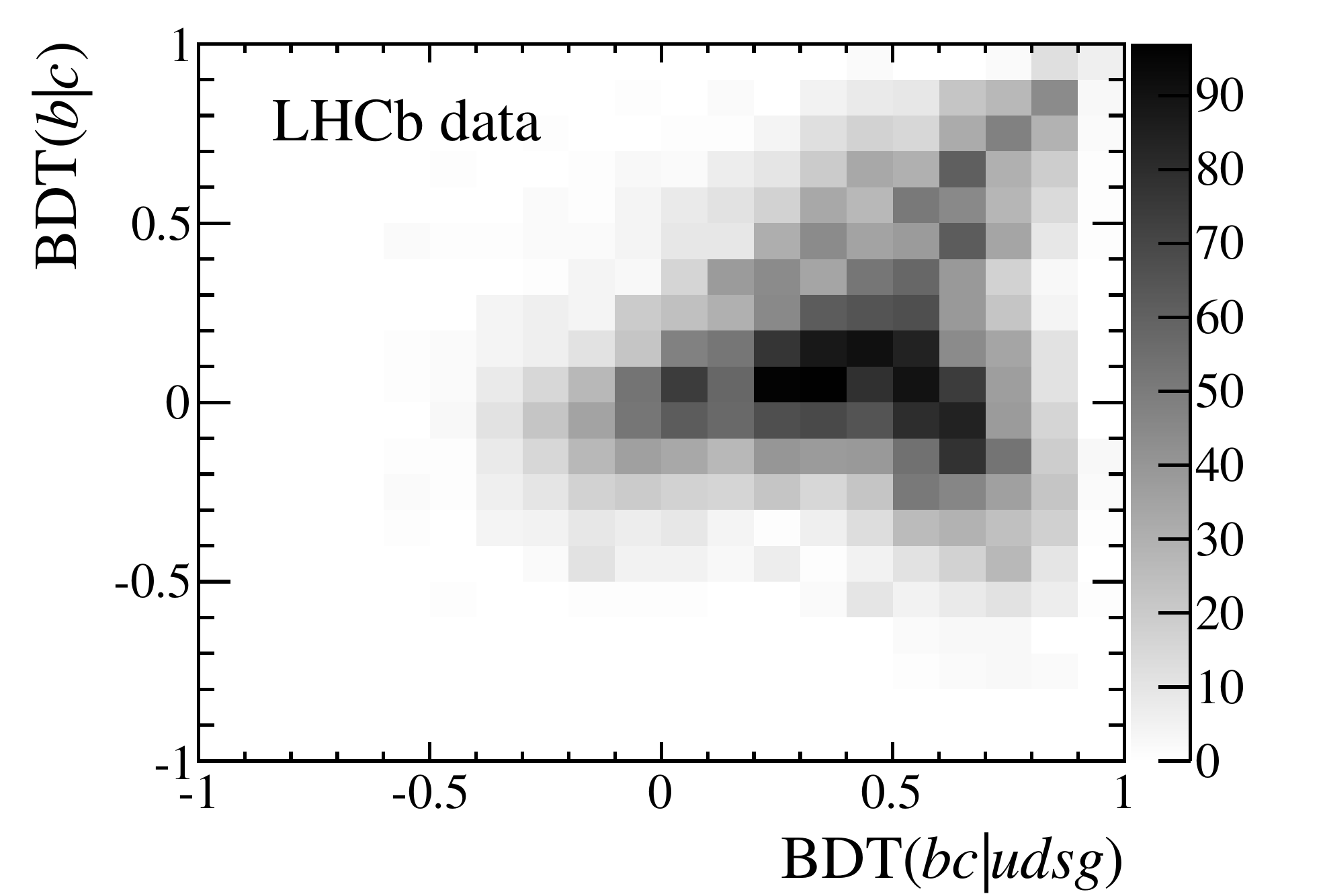}
\includegraphics[width=0.49\textwidth]{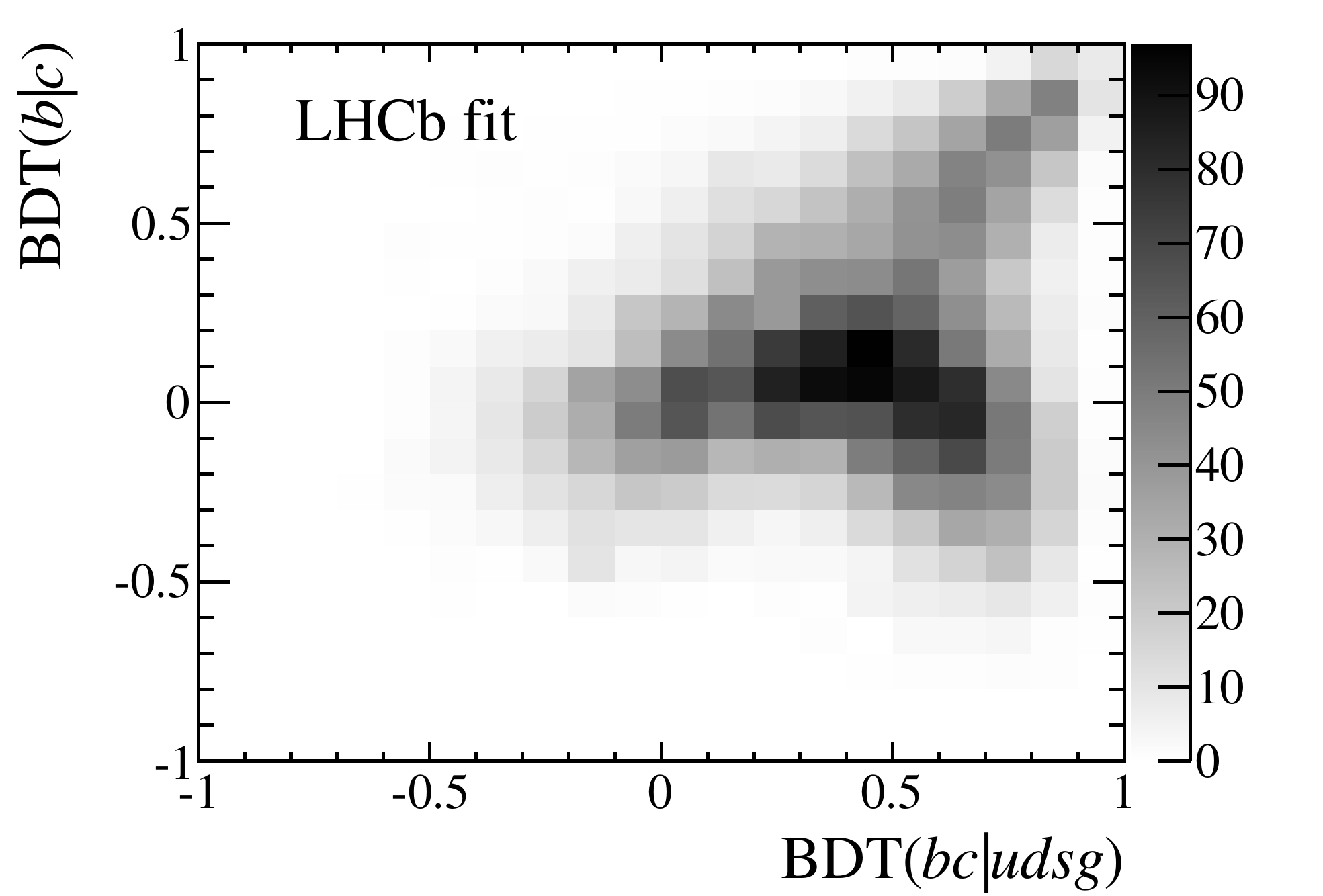}
\includegraphics[width=0.49\textwidth]{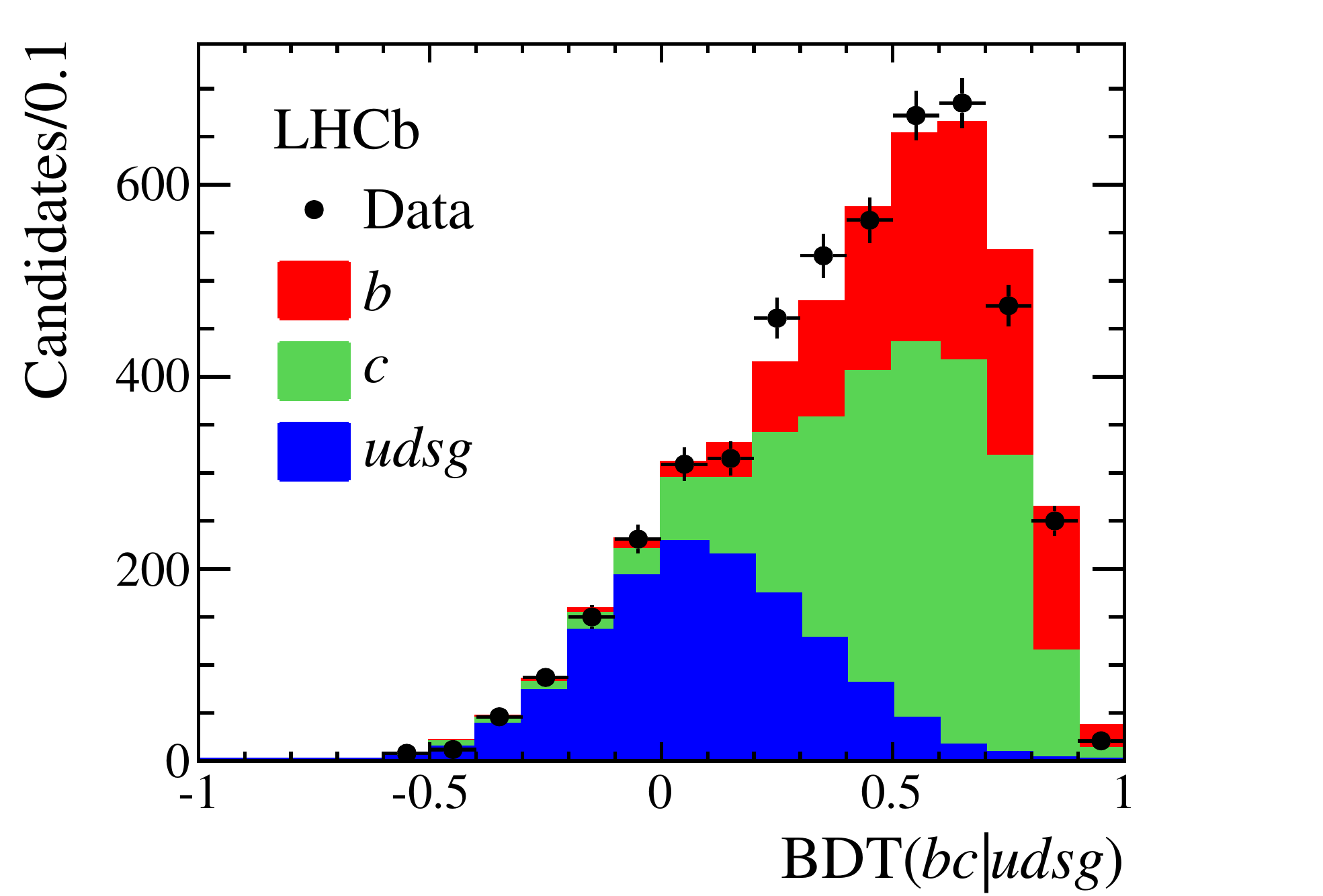}
\includegraphics[width=0.49\textwidth]{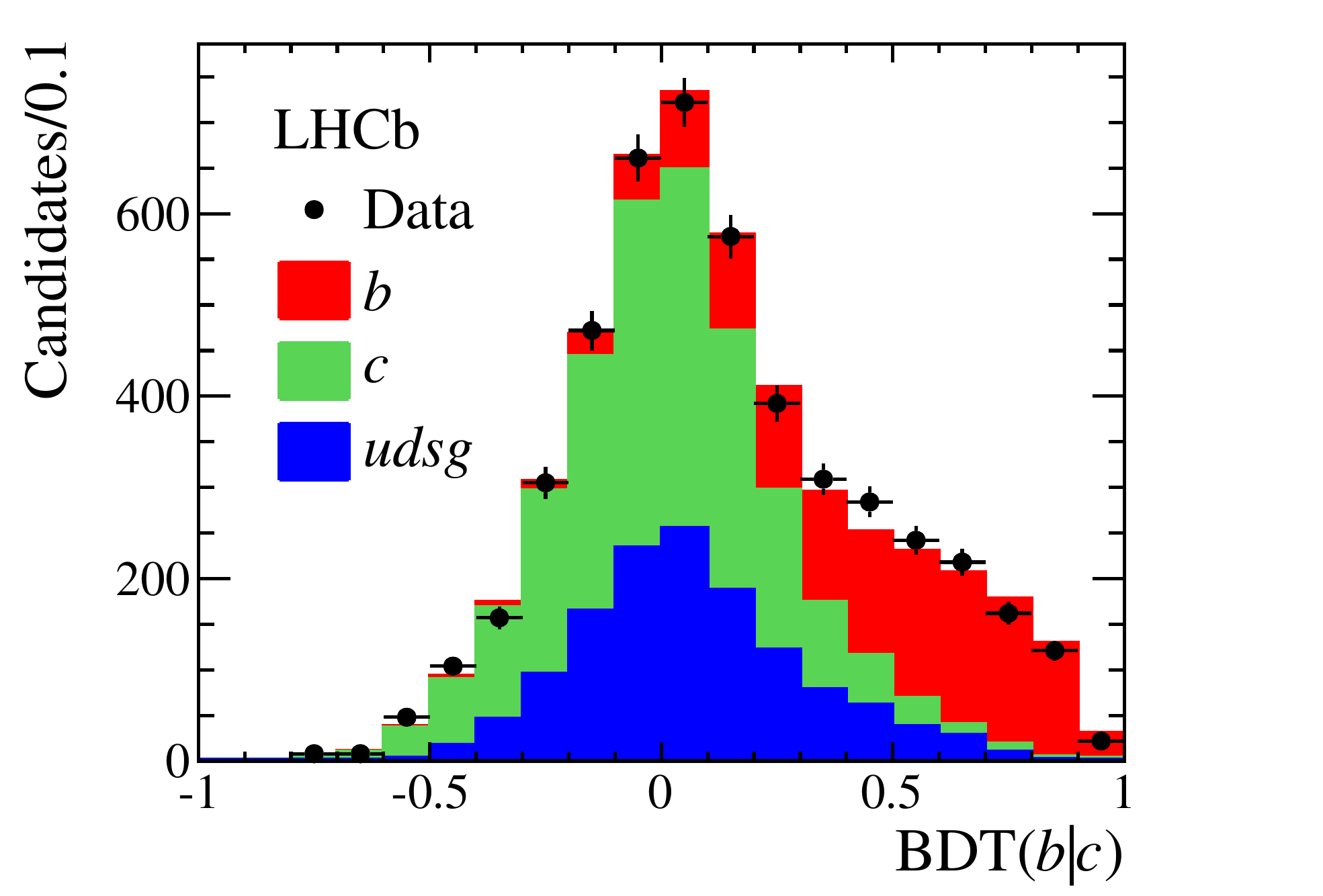}
\includegraphics[width=0.49\textwidth]{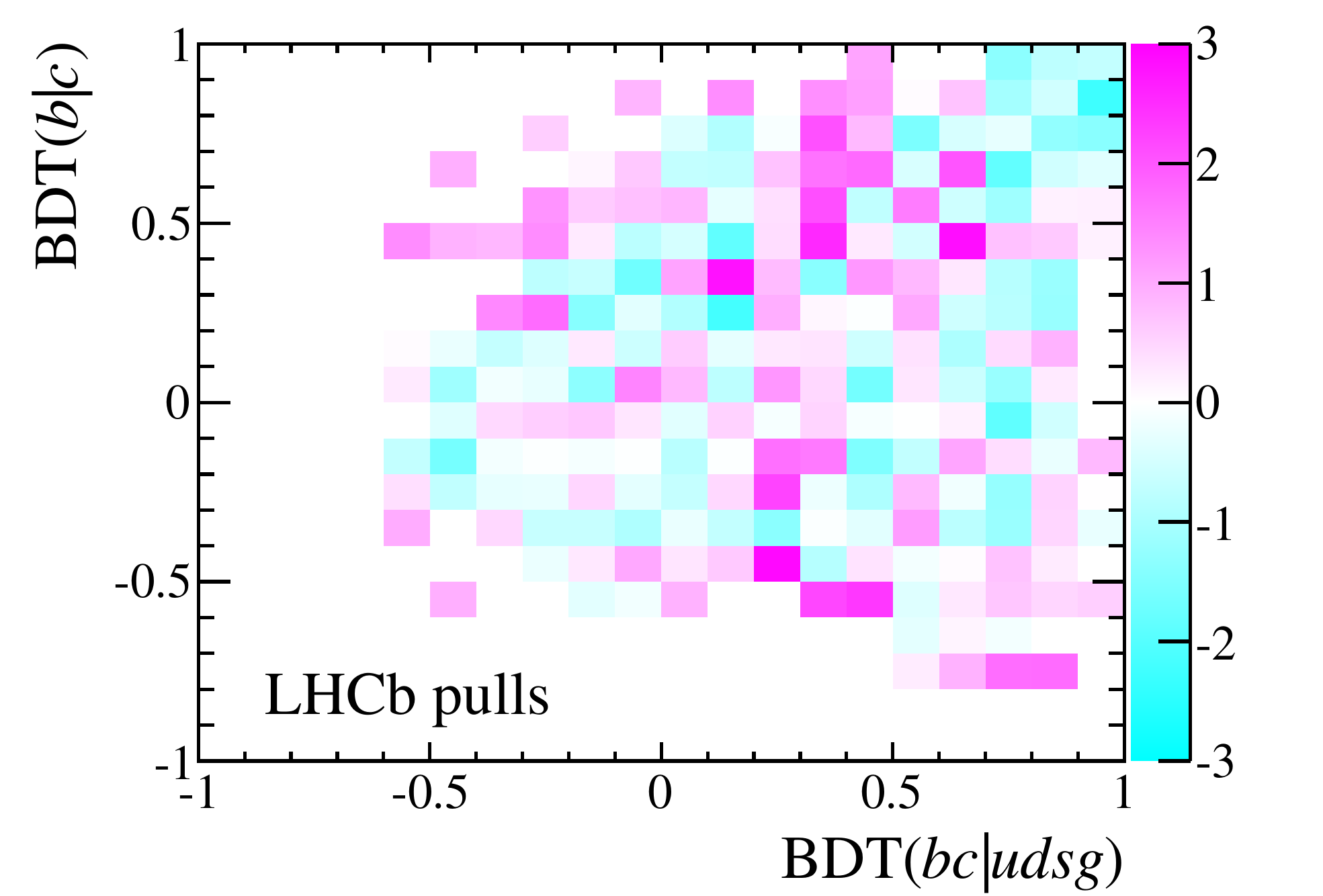}
    \caption{(top left) Two-dimensional SV-tag BDT distribution and (top right) fit for events in the subsample with $\muptr > 0.9$, projected onto the (bottom left) \bdtbcl and (bottom right) \bdtbc axes.  Combined data for $\sqrt{s}=7$ and 8~TeV for both muon charges are shown. \label{fig:svbdt_90_100}}
  \end{center}
\end{figure}

\begin{figure}[]
  \begin{center}
\includegraphics[width=0.49\textwidth]{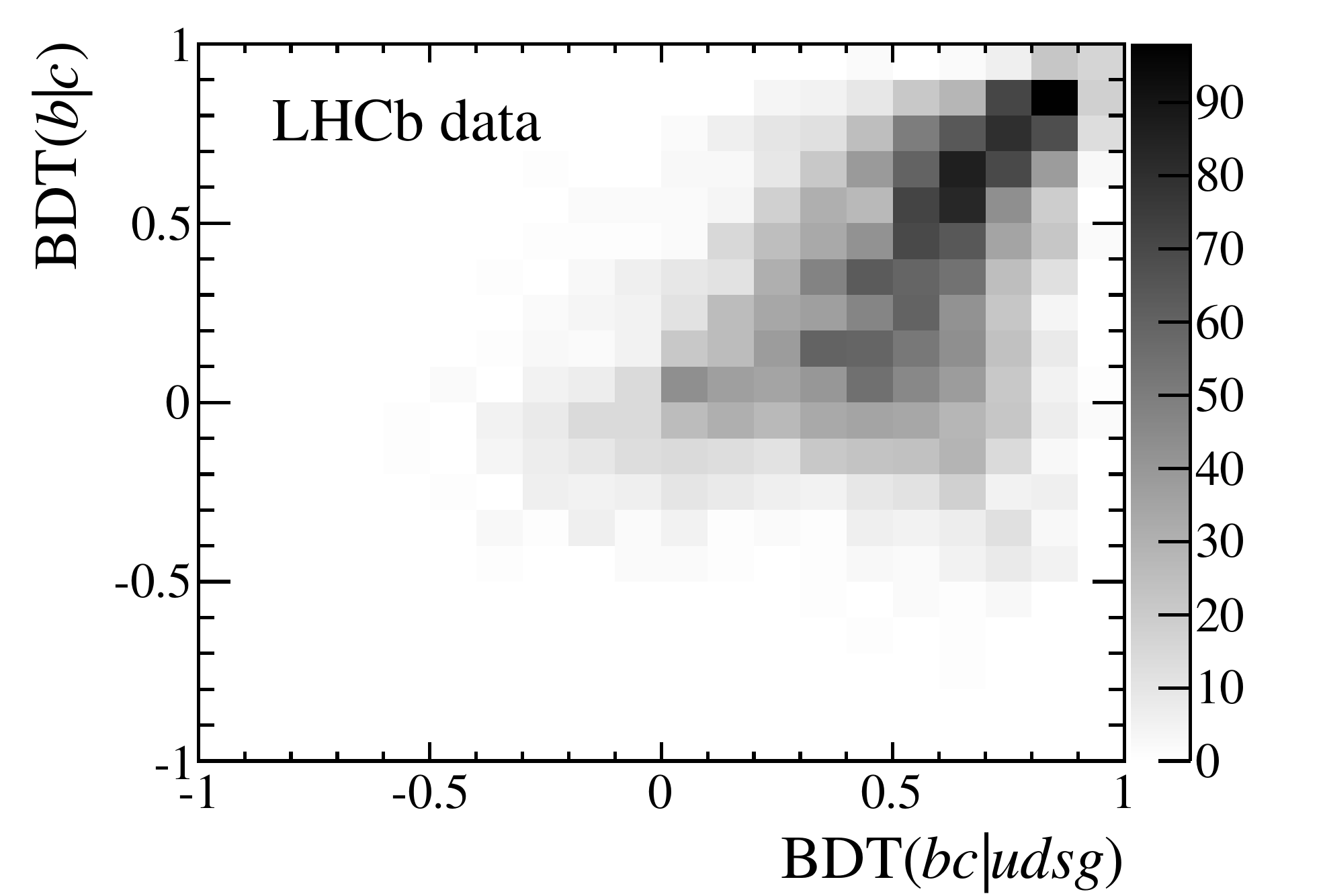}
\includegraphics[width=0.49\textwidth]{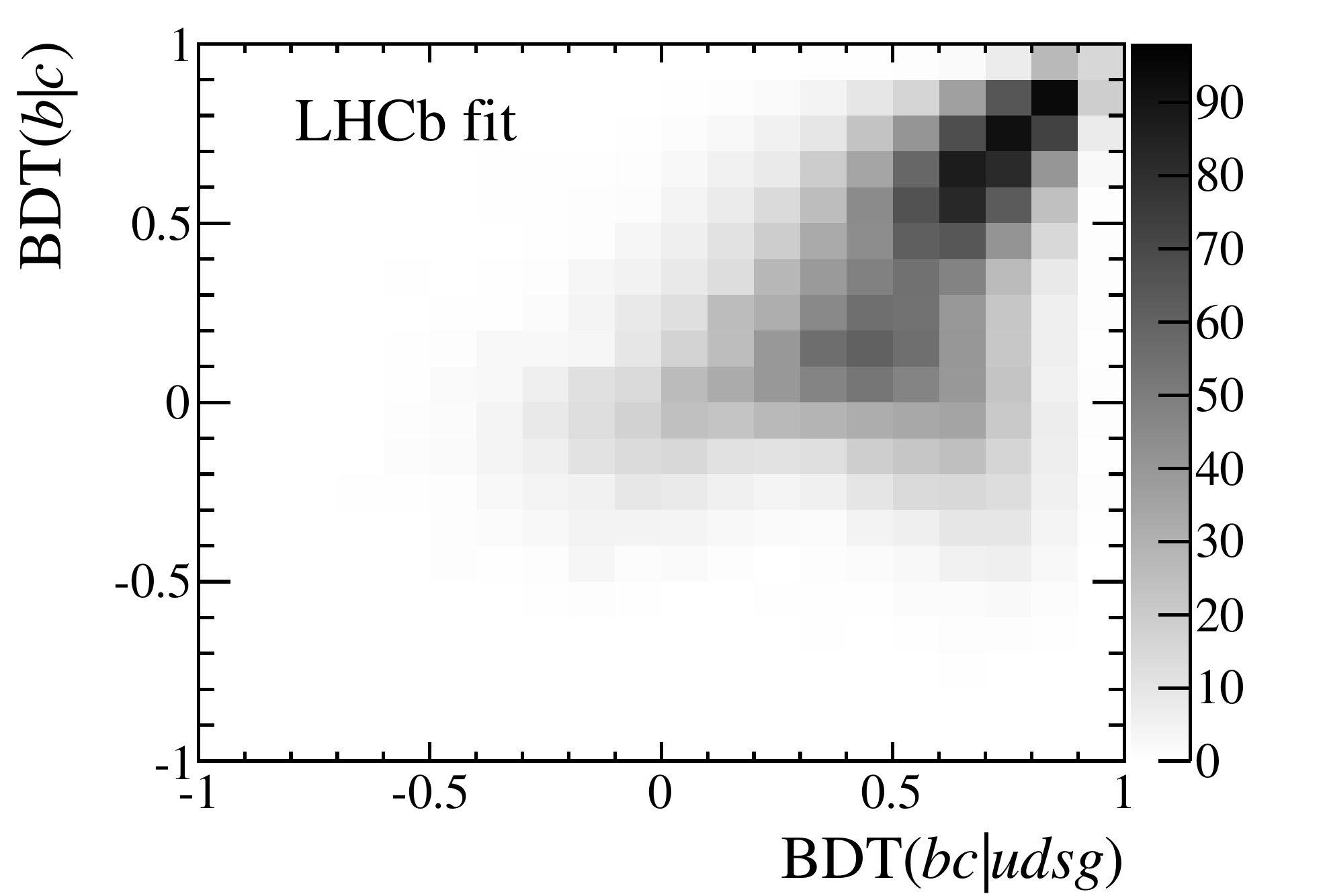}
\includegraphics[width=0.49\textwidth]{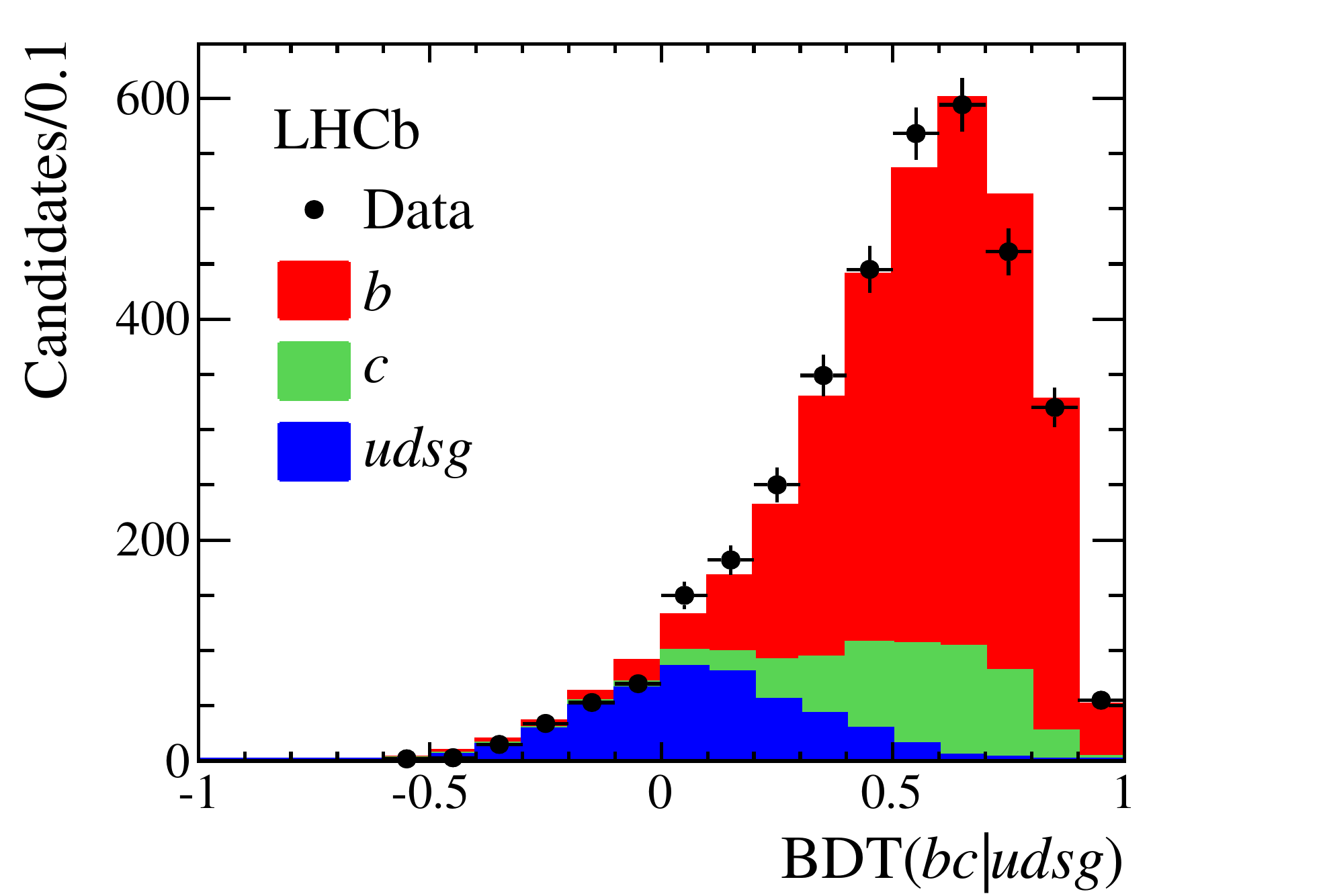}
\includegraphics[width=0.49\textwidth]{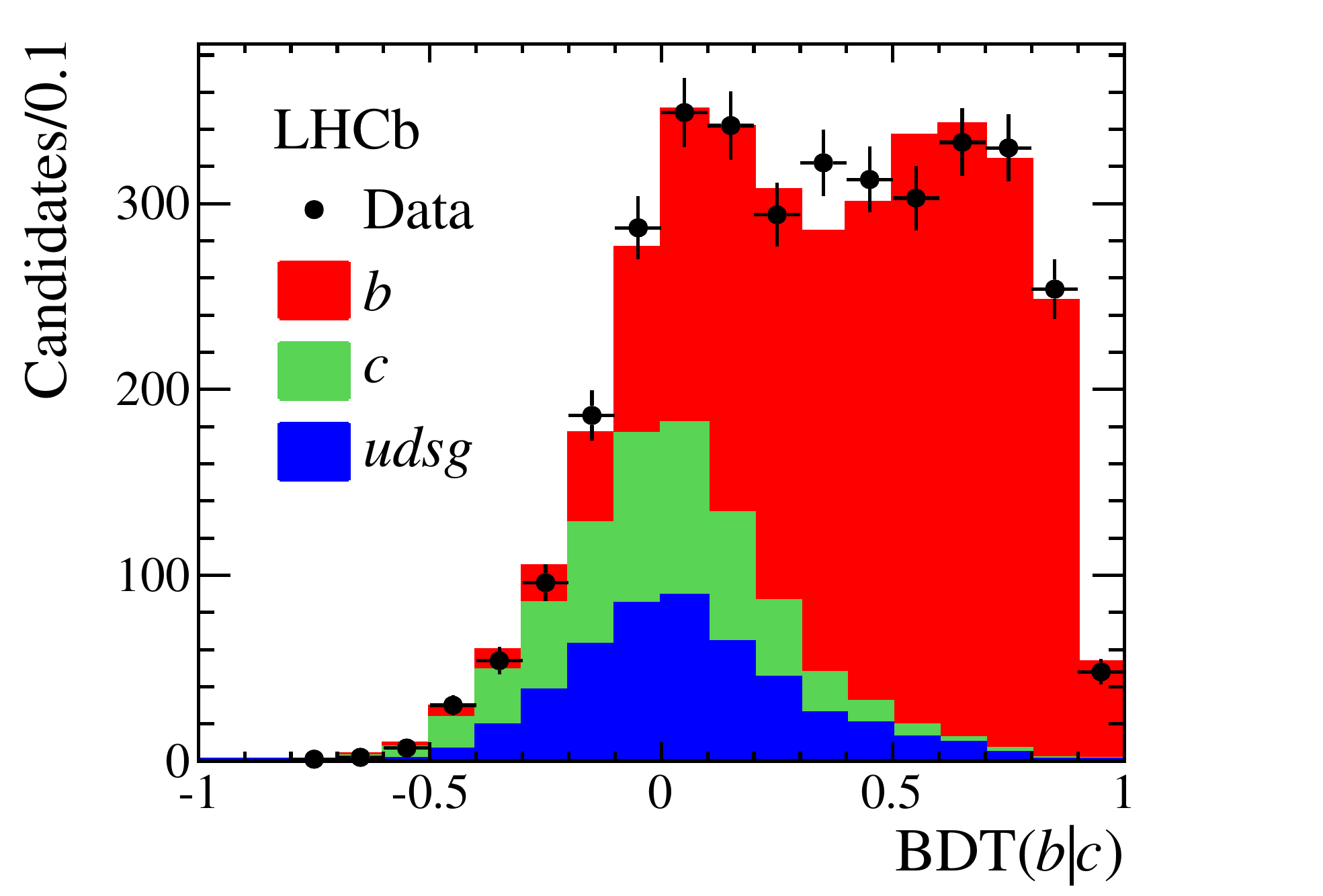}
\includegraphics[width=0.49\textwidth]{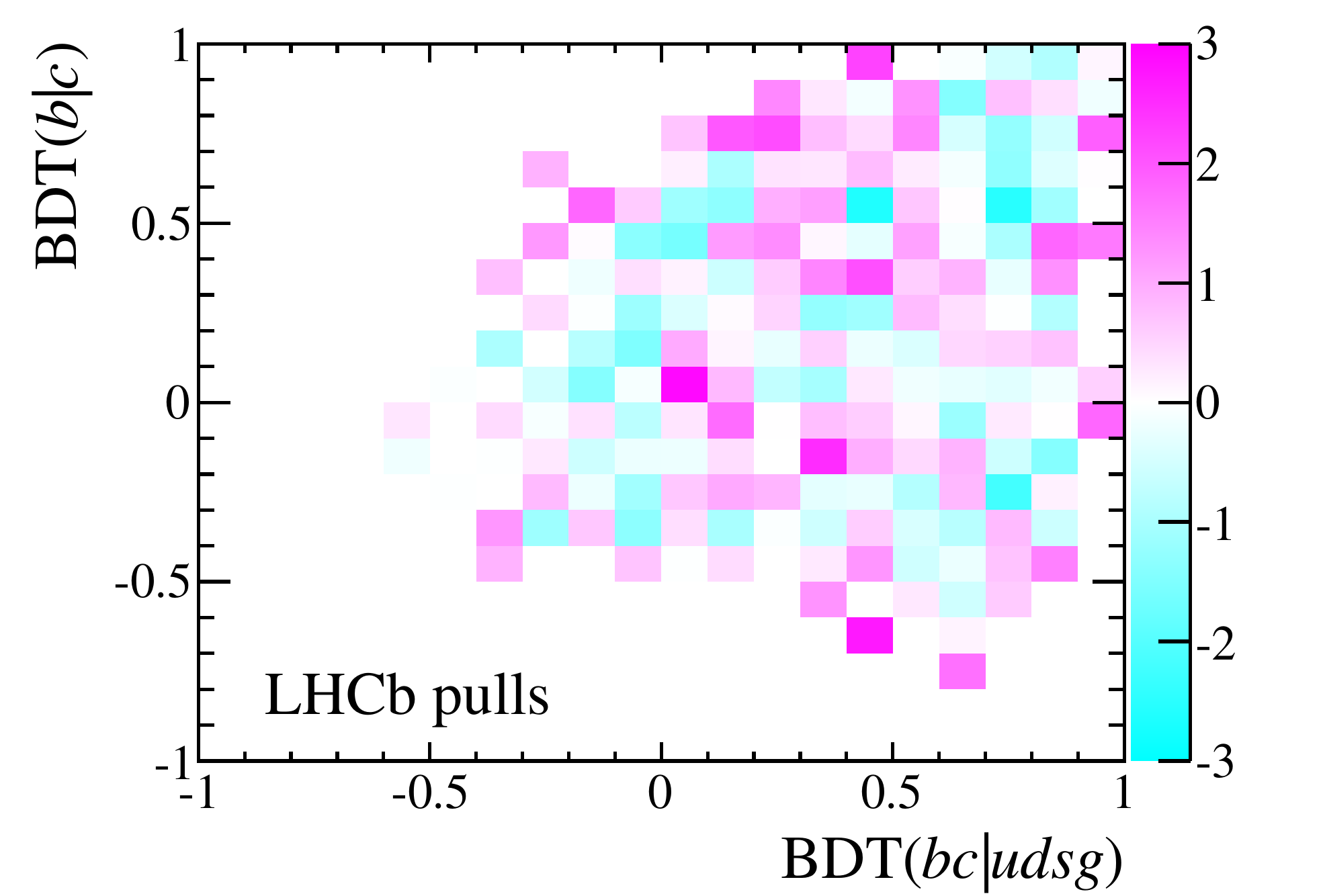}
    \caption{(top left) Two-dimensional SV-tag BDT distribution and (top right) fit for events in the subsample with $0.5 < \muptr < 0.6$, projected onto the (bottom left) \bdtbcl and (bottom right) \bdtbc axes.
Combined data for $\sqrt{s}=7$ and 8~TeV for both muon charges are shown.
\label{fig:svbdt_50_60}}
  \end{center}
\end{figure}

The yields of events with $W$ bosons associated with $b$-tagged and $c$-tagged jets are obtained by fitting the two-dimensional SV-tagger BDT-response distributions for $\sqrt{s}=7$ and 8\tev and for each muon charge separately in bins of \muptr. 
The SV-tagger BDT templates used in this analysis are obtained from the data samples enriched in $b$ and $c$ jets used in Ref.~\cite{LHCb-PAPER-2015-016}. As a consistency check, the two-dimensional BDT distributions are fitted using templates from simulation; the yields shift only by a few percent.
Figure~\ref{fig:svbdt_90_100} shows the BDT distributions combining all data in the most sensitive region, $W+$jet events with $\muptr > 0.9$.  This is the region where the muon carries a large fraction of the muon-jet momentum and is, therefore, highly isolated.  Figure~\ref{fig:svbdt_50_60} shows the distributions in a dijet dominated region ($0.5 < \muptr < 0.6$). In the dijet region the majority of SV-tagged jets associated with the high-\pt muon candidate are found to be $b$ jets. This is due to the large semileptonic branching fraction of $b$ hadrons. In the $W+$jet signal region there are significant contributions from both $b$ and $c$ jets.    

\begin{figure}
  \begin{center}
\includegraphics[width=0.49\textwidth]{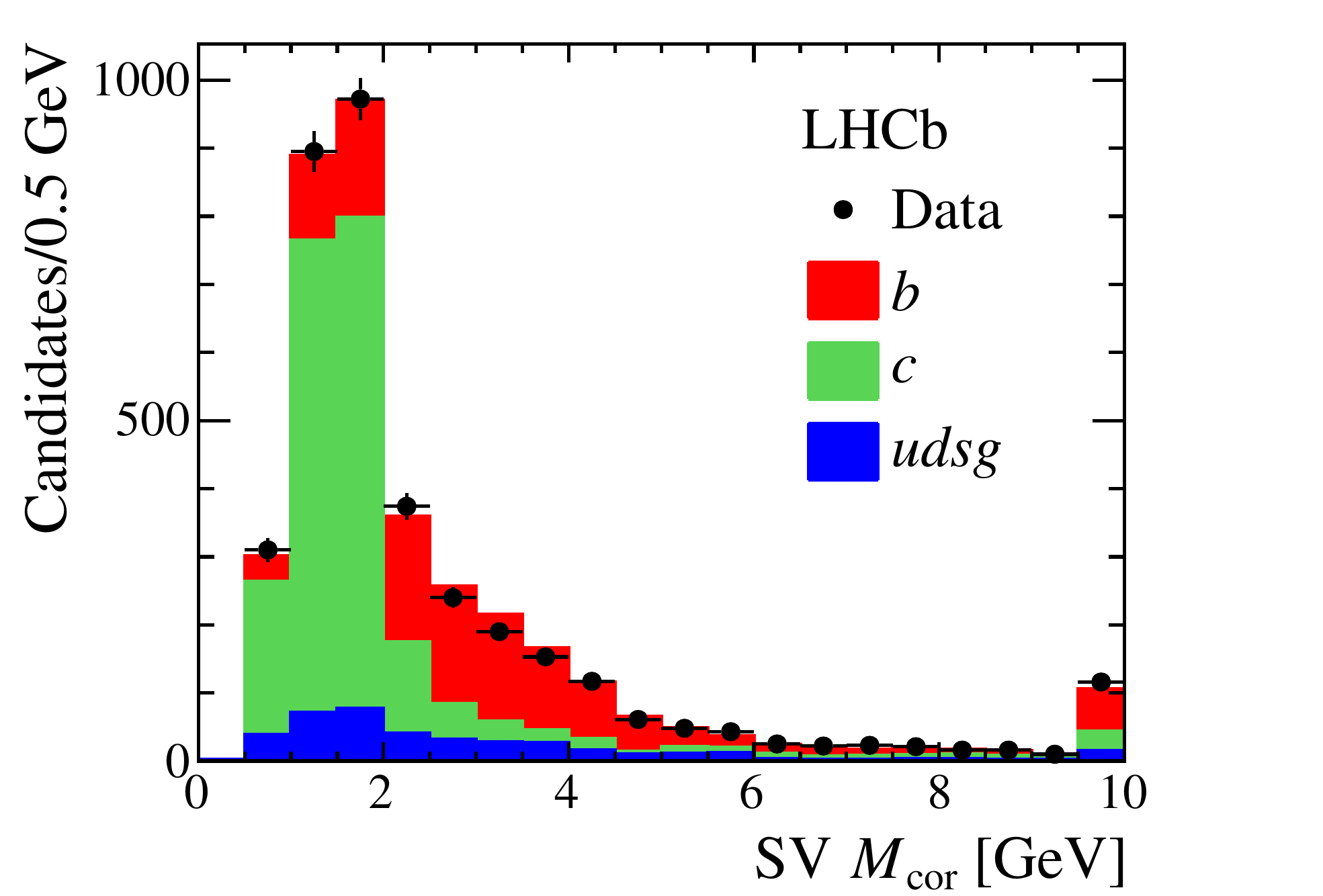}
\includegraphics[width=0.49\textwidth]{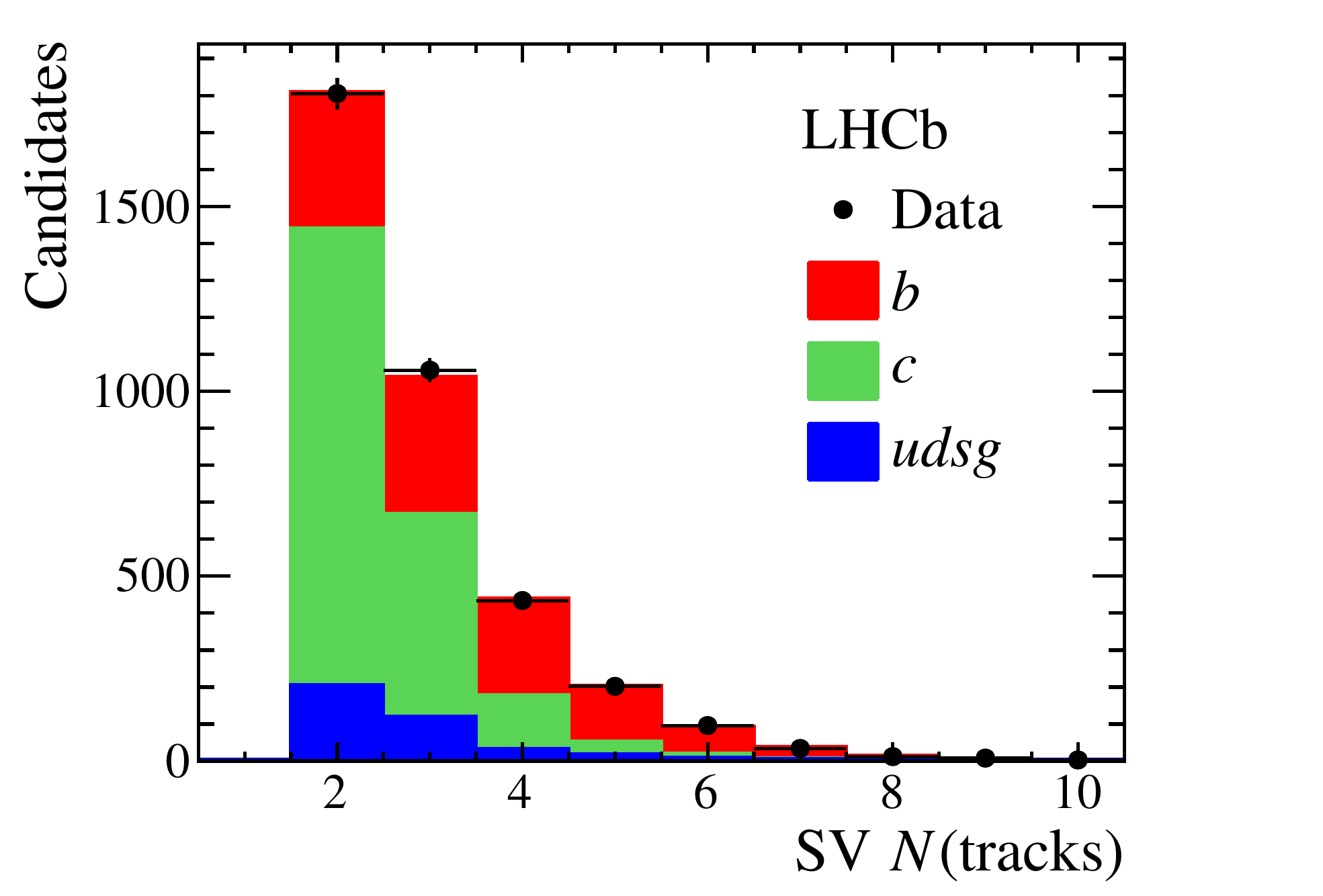}
    \caption{\label{fig:svmcn_90_100} Projections of simultaneous fits of (left) $M_{\rm cor}$ and (right) SV track multiplicity for the SV-tagged subsample with $\bdtbcl > 0.2$ and $\muptr > 0.9$.  
The highest $M_{\rm cor}$ bin includes candidates with $M_{\rm cor} > 10\gev$.
Combined data for $\sqrt{s}=7$ and 8~TeV for both muon charges are shown.
}
  \end{center}
\end{figure}

As a consistency check, the $b$, $c$, and \light yields are obtained in the $\muptr > 0.9$ signal region from a fit using only two of the BDT inputs, both of which rely only on basic SV properties, the track multiplicity and the corrected mass, which is defined as
\begin{equation}
\label{eq:mcor}
M_{\rm cor} = \sqrt{M^2 + |\vec{p}|^2\sin^2{\theta}} + |\vec{p}|\sin{\theta},
\end{equation}
where $M$ and $\vec{p}$ are the invariant mass and momentum of the particles that form the SV, and $\theta$ is the angle between $\vec{p}$ and the flight direction. The corrected mass, which is the minimum mass for a long-lived hadron whose trajectory is consistent with the flight direction, peaks near the $D$ meson mass for $c$ jets and consequently provides excellent discrimination against other jet types. The SV track multiplicity identifies $b$ jets well, since $b$-hadron decays typically produce many displaced tracks. 
In Fig.~\ref{fig:svmcn_90_100}, the distributions of $M_{\rm cor}$ and SV track multiplicity  for a subsample of SV-tagged events with $\bdtbcl > 0.2$ (see Fig.~\ref{fig:svbdt_90_100}) are fitted simultaneously.  
The templates used in these fits are obtained from data in the same manner as the SV-tagger BDT templates.
After correcting for the efficiency of requiring $\bdtbcl > 0.2$, the $b$ and $c$ yields determined from the fits to $M_{\rm cor}$ and SV track multiplicity and from the two-dimensional BDT fits are consistent.
The  mistag probability for $W+$\light events in this sample is found to be approximately $0.3\%$, which agrees with the value obtained from simulation.

\begin{figure}
    \includegraphics[width=0.95\textwidth]{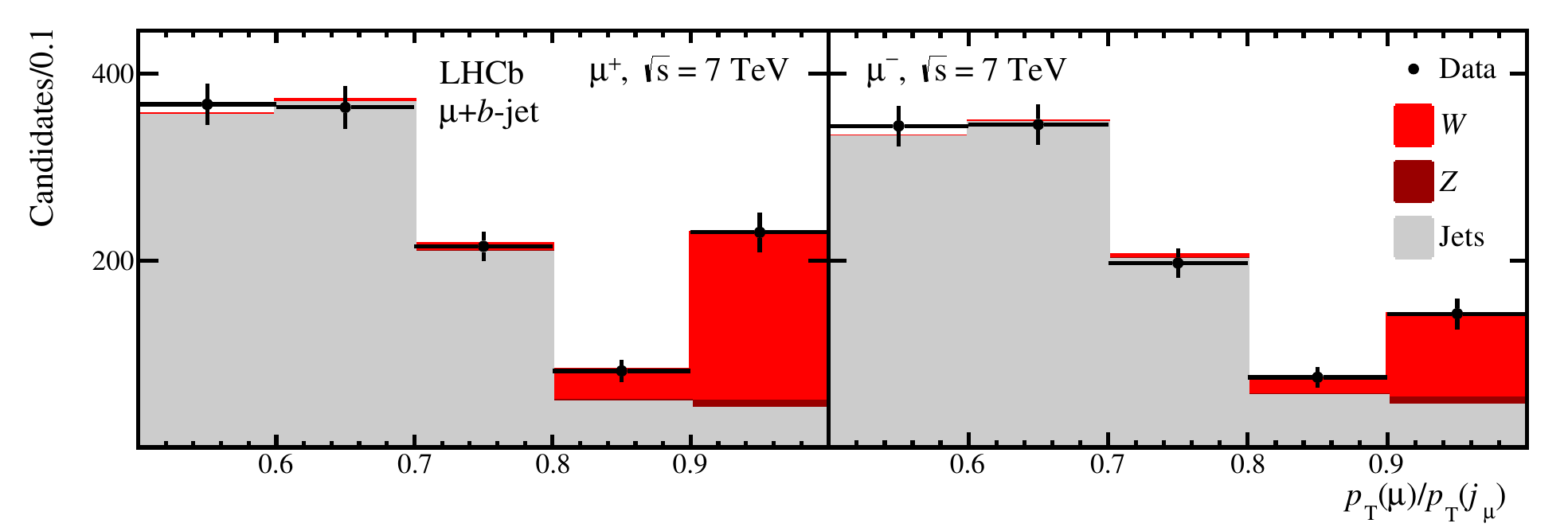}
    \vspace{-0.33in}\\
    \includegraphics[width=0.95\textwidth]{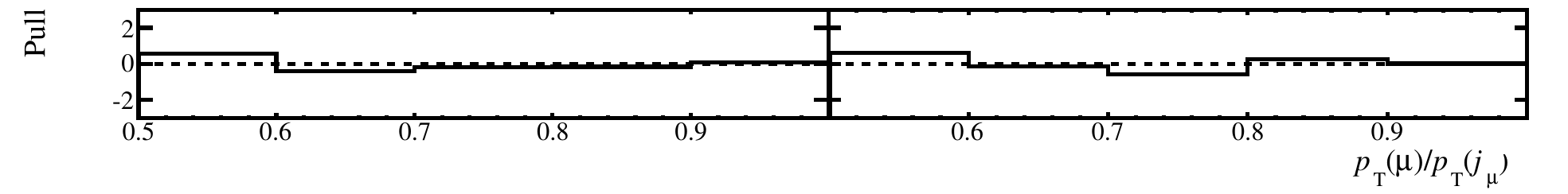}
    \includegraphics[width=0.95\textwidth]{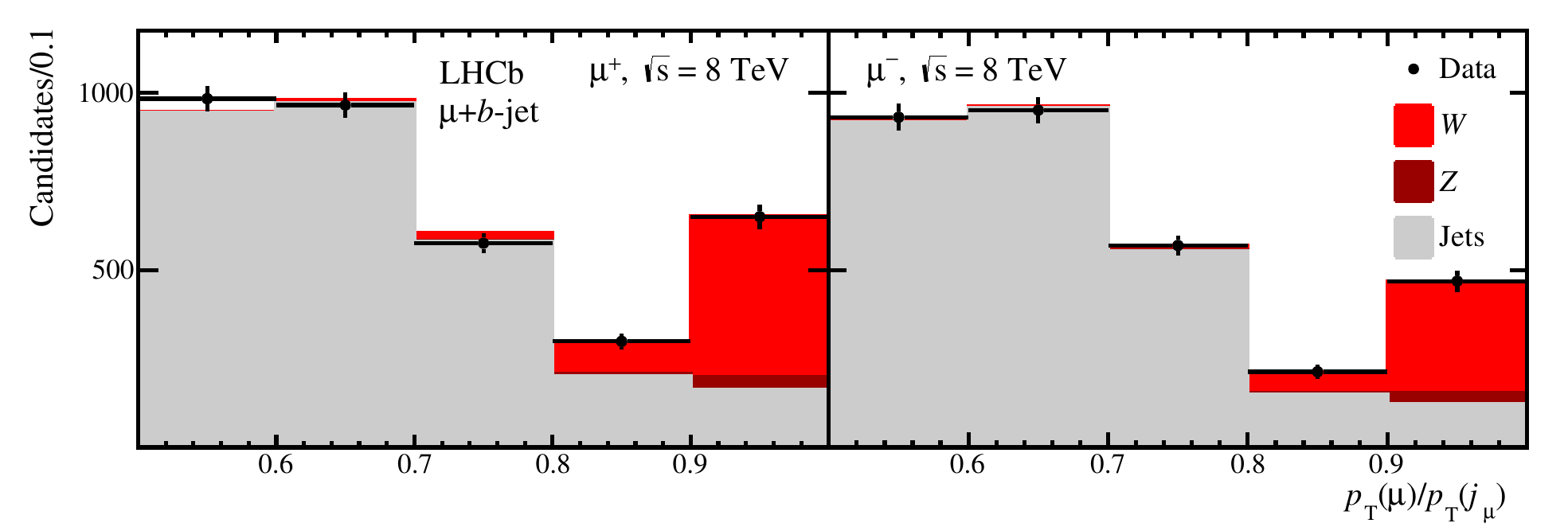}
    \vspace{-0.33in}\\
    \includegraphics[width=0.95\textwidth]{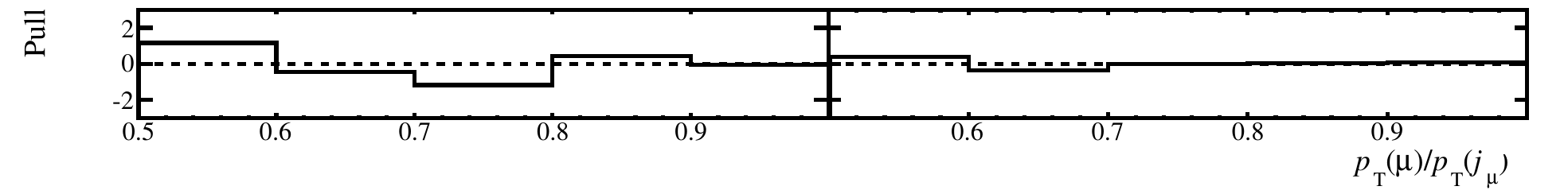}
    \caption{\label{fig:wmuptr_b} Fits to \muptr distributions for $b$-tagged data samples for $\sqrt{s} = 7$ and 8~TeV.}
\end{figure}

\begin{figure}
    \includegraphics[width=0.95\textwidth]{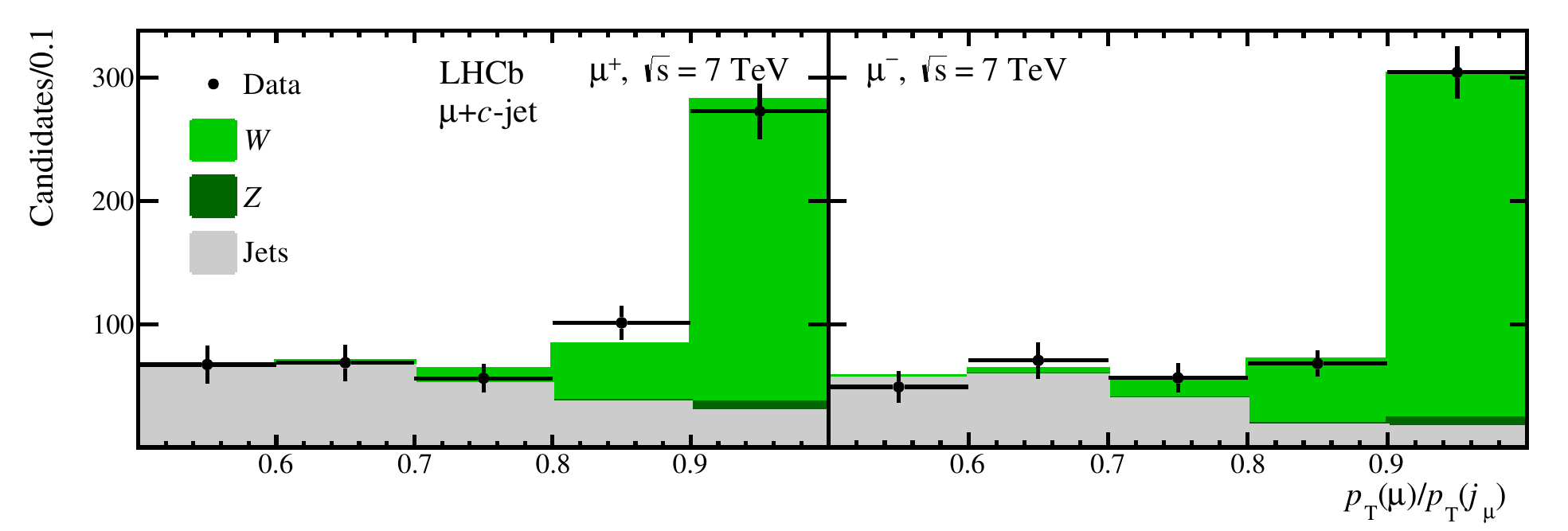}
    \vspace{-0.33in}\\
    \includegraphics[width=0.95\textwidth]{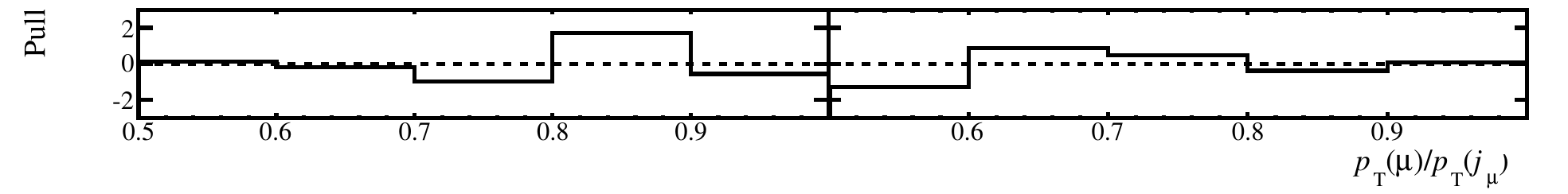}\\
    \includegraphics[width=0.95\textwidth]{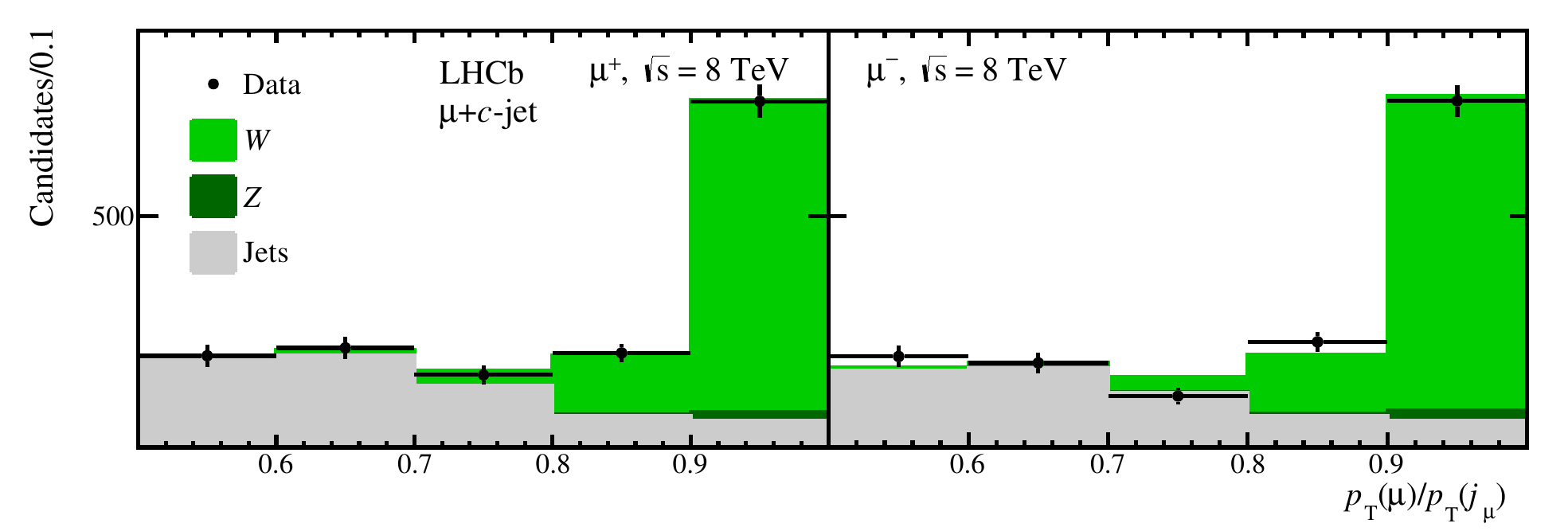}
    \vspace{-0.33in}\\
    \includegraphics[width=0.95\textwidth]{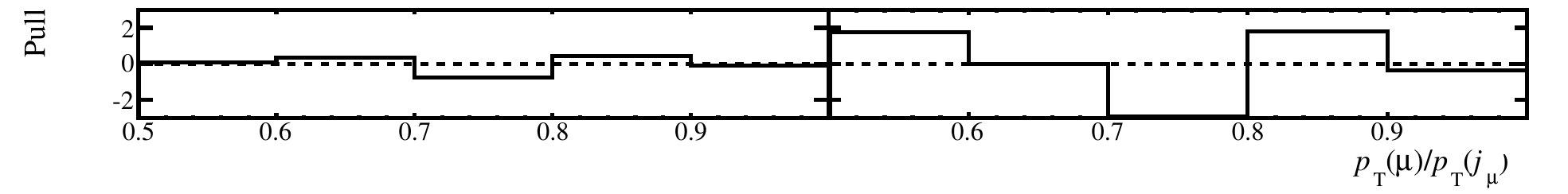}
    \caption{\label{fig:wmuptr_c} Fits to \muptr distributions for $c$-tagged data samples for $\sqrt{s} = 7$ and 8~TeV.}
\end{figure}

From the SV-tagger BDT fits, the $b$ and $c$ yields are obtained in bins of $\sqrt{s}$, muon charge, and \muptr. The \muptr distributions for muons associated with $b$-tagged and $c$-tagged jets are shown in Figs.~\ref{fig:wmuptr_b} and \ref{fig:wmuptr_c}. These distributions are fitted to determine the \wb and \wc final-state yields as in the inclusive $W+$jet sample. The $Z\!+\!b$ and $Z\!+\!c$ yields are obtained by fitting the SV-tagger BDT distributions in the fully reconstructed $Z+$jet data samples and then correcting for the missed-muon probability.
The fits are shown in Figs.~\ref{fig:wmuptr_b} and \ref{fig:wmuptr_c} for each muon charge and center-of-mass energy. 
The yields obtained still include contributions from top quark production and $Z\to\tau\tau$.

\begin{figure}
  \begin{center}
    \includegraphics[width=0.49\textwidth]{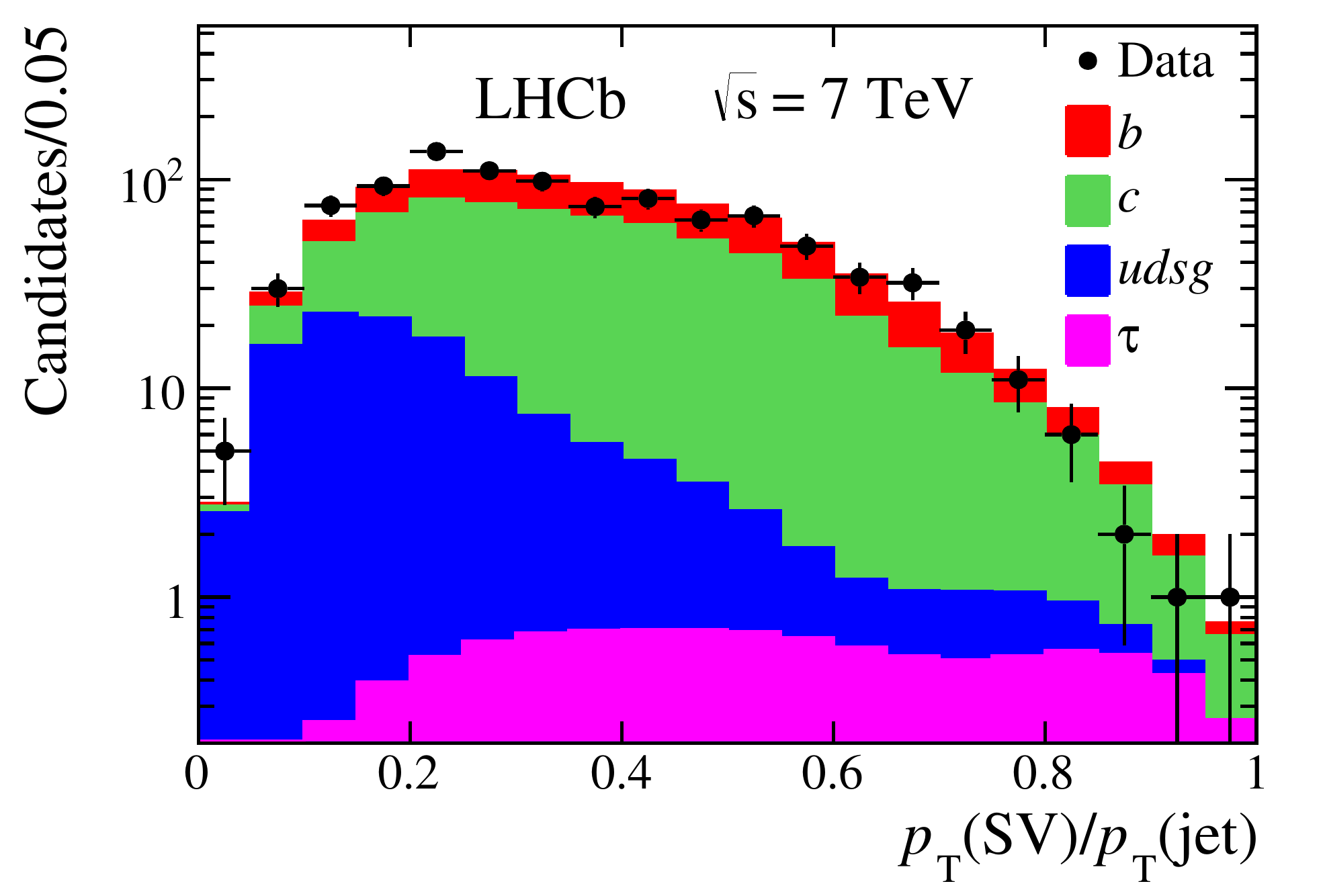}
    \includegraphics[width=0.49\textwidth]{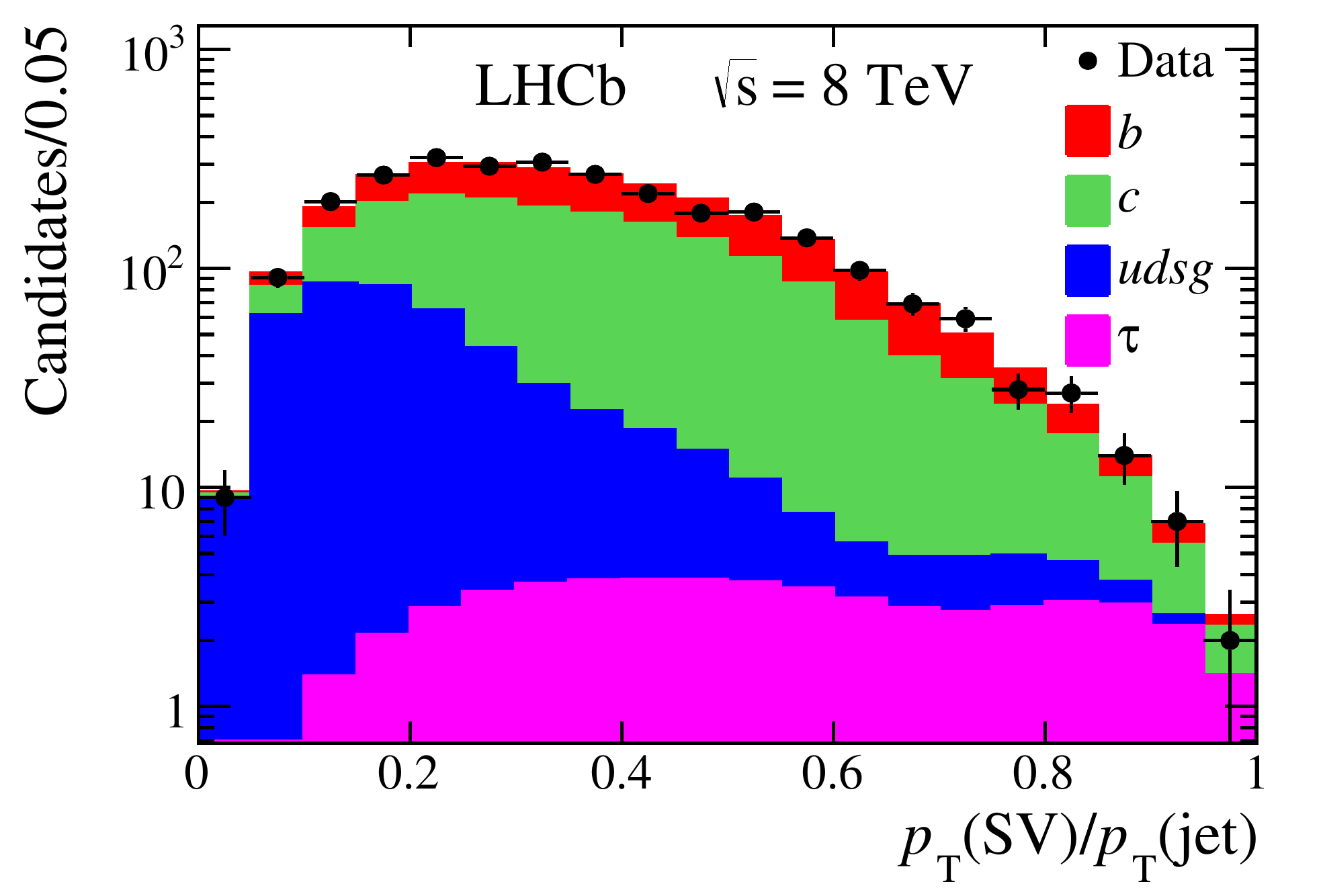}
    \vspace{-0.35in}\\
    \hspace{-0.11in}
    \includegraphics[width=0.49\textwidth]{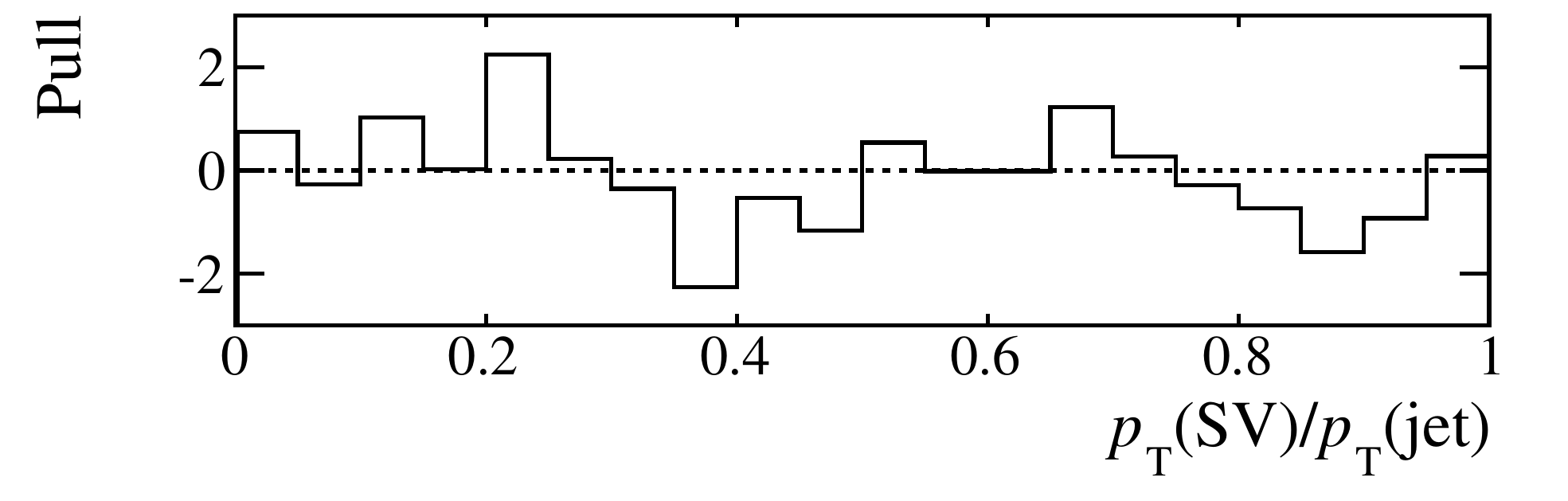}
    \includegraphics[width=0.49\textwidth]{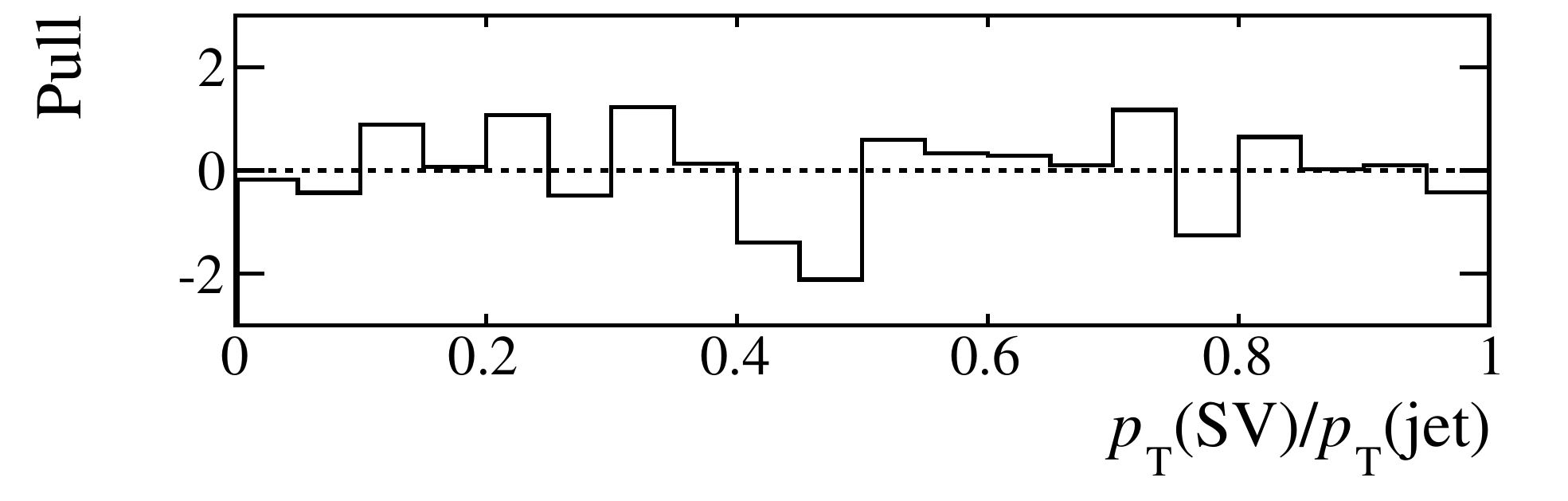}
    \caption{\label{fig:tau_fit}Fits to the  $\pt(\mathrm{SV})/\pt(j)$ distributions in (left) $7\protect\tev$ and (right) $8\protect\tev$ data for candidates with $\muptr > 0.9$ and $\bdtbcl > 0.2$.}
  \end{center}
\end{figure}

The $Z\to\tau\tau$ background, where one $\tau$ lepton decays into a muon and the other into a hadronic jet, contaminates the \wc sample due to the similarity of the $c$-hadron and $\tau$ lepton masses. The $\pt({\rm SV})/\pt(j)$ distribution, where $\pt({\rm SV})$ is the transverse momentum of the particles that form the SV, is used to discriminate between $c$ and $\tau$ jets, since SVs produced from $\tau$ decays usually carry a larger fraction of the jet energy than SVs from $c$-hadron decays. Figure~\ref{fig:tau_fit} shows fits to the $\pt({\rm SV})/\pt(j)$ distributions observed in data where the $b$ and \light yields are fixed using the results of BDT fits performed on the data samples.  A requirement of $\bdtbcl > 0.2$ is applied to this sample to remove the majority of SV-tagged \light jets while retaining 90\% of $b$, $c$ and $\tau$ jets.
The only free parameter in these fits is the fraction of jets identified as charm in the SV-tagger BDT fits that originate from $\tau$ leptons.  The $\pt({\rm SV})/\pt(j)$ templates are obtained from simulation.
The $Z\to\tau\tau$ yields are consistent with SM expectations and are about 25 times smaller than the \wc yields.  
These results are extrapolated to the inclusive sample using simulation.

The top quark background is determined in the dedicated analysis of Ref.~\cite{TOP}, where a reduced fiducial region is used to enrich the relative top quark content. The yields and charge asymmetries of the \wb final state as functions of $\pt(\mu+b)$ are used to discriminate between \wb and top quark production. The results obtained in Ref.~\cite{TOP} are consistent with SM expectations and are extrapolated to the fiducial region of this analysis using simulation based on NLO calculations.  
The extrapolated top quark yields are subtracted from the observed number of \wb candidates to obtain the signal yields.  
Top quark production is found to be responsible for about 1/3 of events that contain a $W$ boson and $b$ jet. 
A summary of all signal yields is given in Table~\ref{tab:yields}.

\begin{table}[]
  \newcommand{\pad}{\phantom{1}}
  \caption{\label{tab:yields} Summary of signal yields. The two $Zj$ yields denote the charge of the muon on which the trigger requirement is made.  The $Zj$ yields given are the numbers of candidates observed, while the $W$ boson yields are obtained from fits.
    The yield due to top quark production is subtracted in these results.
}
  \begin{center}
    \begin{tabular}{lrrrr}
      \toprule
      & \multicolumn{2}{c}{$7\tev$}  & \multicolumn{2}{c}{$8\tev$} \\
      \multicolumn{1}{l}{Mode} & \multicolumn{1}{c}{$\mu^+$} & \multicolumn{1}{c}{$\mu^-$} & \multicolumn{1}{c}{$\mu^+$} & \multicolumn{1}{c}{$\mu^-$} \\
      \midrule
      $Zj$  & \multicolumn{1}{c}{2364} & \multicolumn{1}{c}{2357} & \multicolumn{1}{c}{6680} & \multicolumn{1}{c}{6633} \\      
      $Wj$ & $27\,400\pm500$ & $17\,500\pm400$ & $70\,700\pm1100$ & $44\,800\pm800$ \\
      $Wb$-tag & $160\pm31$\pad & $51\pm27$\pad & $400\pm43$\pad\pad & $236\pm45$\pad \\
      $Wc$-tag & $295\pm36$\pad & $338\pm31$\pad & $795\pm56$\pad\pad & $802\pm55$\pad  \\
      \bottomrule
    \end{tabular}
  \end{center}
\end{table}

\section{Systematic uncertainties}\label{sec:uncertainty}

A summary of the relative systematic uncertainties separated by source for each measurement is provided in Table~\ref{tab:sys}. A detailed description of each contribution is given below.

The \pt distributions of muons from $W$ and $Z$ bosons produced in association with $b$, $c$ and \light jets are nearly identical. This results in a negligible uncertainty from muon trigger and reconstruction efficiency on cross section ratios involving only $W$ bosons. In the ratios $\sigma(W^+j)/\sigma(Zj)$ and $\sigma(W^-j)/\sigma(Zj)$, the muon from the $Z$ boson decay with the same charge as that from the $W$ decay is required to satisfy the same trigger and selection requirements as the $W$ boson muon, giving negligible uncertainty from the trigger and selection efficiency. 
The efficiency for reconstructing and selecting the additional muon from the $Z$ boson decay is obtained from the data-driven studies of Ref.~\cite{LHCb-PAPER-2015-001}. A further data-driven correction is applied to account for the higher occupancy in events with jets~\cite{LHCb-PAPER-2013-058}; a 2\% systematic uncertainty is assigned to this correction.

The GEC efficiency is obtained following Ref.~\cite{LHCb-PAPER-2013-058}: an alternative dimuon trigger requirement with a looser GEC is used to determine the fraction of events that are rejected. The GEC efficiencies for all final states are found to be consistent within a statistical precision of 1\%, which is assigned as a systematic uncertainty. As a further check, the number of jets per event reconstructed in association with $W$ or $Z$ bosons is compared and found to be consistent.

The jet reconstruction efficiencies for heavy-flavor and \light jets in simulation are found to be consistent within $2\%$, which is assigned as a systematic uncertainty for flavor-dependencies in the jet-reconstruction efficiency.  The jet \pt detector response is studied with a data sample enriched in $b$ jets using SV tagging. The $\pt({\rm SV})/\pt(j)$ distribution observed in data is compared to templates obtained from simulation in bins of jet \pt. The resolution and scale in simulation for each jet \pt bin are varied to find the best description of the data and to construct a data-driven unfolding matrix. 
The results obtained using this unfolding matrix are consistent with those obtained using a matrix determined by studies of \pt balance in $Z+$jet events~\cite{LHCb-PAPER-2013-058}, where no heavy-flavor tagging is applied.
The unfolding corrections are at the percent level and their statistical precision is assigned as the uncertainty.  

The heavy-flavor tagging efficiencies are measured from data in Ref.~\cite{LHCb-PAPER-2015-016}, where a 10\% uncertainty is assigned for $b$ and $c$ jets. The cross-check fits of Sect.~\ref{sec:yields}, using the corrected mass and track multiplicity, remove information associated with jet quantities, such as \pt, from the yield determination and produce yields consistent at the 5\% level. This is assigned as the uncertainty for the SV-tagged yield determination.

The $W$ boson template for the \muptr distribution is derived from data, as described in Sect.~\ref{sec:yields}. The fit is repeated using variations of this template, {\em e.g.} using a template taken directly from simulation and using separate templates for $W^+$ and $W^-$, to assess a systematic uncertainty. The dijet templates are obtained from data in a dijet-enriched region.  The residual, small $W$ boson contamination is subtracted using two methods: the $W$ boson yield expected in the dijet-enriched region is taken from simulation; and the \muptr distribution in the dijet-enriched region is fitted to a parametric function to estimate the $W$ boson yield. The difference in the $W$ boson yields obtained using these two sets of dijet templates is at most 2\%.  
The uncertainty on $W/Z$ ratios due to the $W$ boson and dijet templates is 4\%.  The uncertainty due to the $W$ boson template cancels to good approximation in the measurements of $\sigma(Wb)/\sigma(Wj)$ and $\sigma(Wc)/\sigma(Wj)$; however, the uncertainty due to the dijet templates is larger due to the enhanced dijet background levels.  Variations of the dijet templates are considered, with 10\% and 5\% uncertainties assigned on $\sigma(Wb)/\sigma(Wj)$ and $\sigma(Wc)/\sigma(Wj)$.

The systematic uncertainty from top quark production is taken from Ref.~\cite{TOP}, while the systematic uncertainty from $Z\to\tau\tau$ is evaluated by fitting the data using variations of the $\pt({\rm SV})/\pt(j)$ templates. All other electroweak backgrounds are found to be negligible from NLO predictions.  All $W\to\mu\nu$ yields have a small contamination from $W\to\tau\to\mu$ decays that cancels in all cross section ratios except for the $W/Z$ ratios. A scaling factor of 0.975, obtained from simulation, is applied to the $W$ boson yields.  A 1\% uncertainty is assigned to the scale factor, which is obtained from the difference between the correction factor from simulation and a data-driven study of this background~\cite{LHCb-PAPER-2014-033} for inclusive $W\to\mu\nu$ production.  

The trigger, reconstruction and selection requirements are consistent with being charge symmetric~\cite{LHCb-PAPER-2014-033}, which results in negligible uncertainty on $\mathcal{A}(Wb)$ and $\mathcal{A}(Wc)$.   
Unfolding of the jet \pt detector response is performed independently for $W^+$ and $W^-$ bosons, with the statistical uncertainties on the corrections to the charge asymmetries assigned as systematic uncertainties.
The uncertainty on the \wb and \wc yields from the BDT templates is included in the charge asymmetry uncertainty due to the fact that the fractional jet content of the SV-tagged samples is charge dependent.  
The uncertainty on the charge asymmetries due to determination of the $W$ boson yields is evaluated using an alternative method for obtaining the charge asymmetries.  The raw charge asymmetry in the $b$-jet and $c$-jet yields in the $\muptr > 0.9$ region is obtained from the SV-tagger BDT fits. The $Z+$jet and dijet backgrounds are charge symmetric at the percent level and contribute at most to 20\% of the events in this \muptr region. Therefore, $\mathcal{A}(Wb)$ and $\mathcal{A}(Wc)$ are approximated by scaling the raw asymmetries by the inverse of the $W$ boson purity in the $\muptr > 0.9$ region. A small correction must also be applied to $\mathcal{A}(Wb)$ to account for top quark production. The difference between the asymmetries from this method and the nominal method is assigned as a systematic uncertainty from $W$ boson signal determination.
The uncertainty on $\mathcal{A}(Wb)$  due to top quark production is taken from Ref.~\cite{TOP}.

\begin{table}
  \begin{center}
    \caption{Systematic uncertainties.  Relative uncertainties are given for cross section ratios and absolute uncertainties for charge asymmetries.\label{tab:sys}}
    \newcommand{\pad}{\phantom{$^\dagger$}}
    \begin{tabular}{lrrrrr}
      \toprule
      \multicolumn{1}{c}{Source} & $\frac{\sigma(Wb)}{\sigma(Wj)}$ & $\frac{\sigma(Wc)}{\sigma(Wj)}$ & $\frac{\sigma(Wj)}{\sigma(Zj)}$ &  $\mathcal{A}(Wb)$ & $\mathcal{A}(Wc)$ \\ 
      \midrule
      Muon trigger and selection & $-$\pad & $-$\pad & 2\% & $-$\pad & $-$\pad \\
      GEC & 1\% & 1\% & 1\% & $-$\pad & $-$\pad \\
      Jet reconstruction & 2\% & 2\% & $-$\pad & $-$\pad & $-$\pad \\
      Jet \pt  & 2\% & 2\% & 1\% & 0.02 & 0.02 \\
      $(b,c)$-tag efficiency & 10\% & 10\% & {} & $-$\pad & $-$\pad \\
      SV-tag BDT templates & 5\% & 5\% & {} & 0.02 & 0.02\\
      \muptr templates & 10\% & 5\% & 4\% & 0.08 & 0.03 \\
      Top quark  & 13\% & $-$\pad & $-$\pad & 0.02 & {} \\
      $Z\to\tau\tau$ & $-$\pad & 3\% & $-$\pad & $-$\pad & $-$\pad \\
      Other electroweak & $-$\pad & $-$\pad & $-$\pad & $-$\pad & $-$\pad \\
      $W\to\tau\to\mu$ & $-$\pad & $-$\pad & 1\% & $-$\pad & $-$\pad \\
      \midrule
      Total & 20\% & 13\% & 5\% & 0.09 & 0.04\\
      \bottomrule
    \end{tabular}
  \end{center}
\end{table}

\begin{table}
  \begin{center}
    \caption{\label{tab:summary}Summary of the results and SM predictions.  For each measurement the first uncertainty is statistical, while the second is systematic.
All results are reported within a fiducial region that requires a jet with $\pt > 20\gev$ in the pseudorapidity range $2.2 < \eta < 4.2$, a muon with $\pt > 20\gev$ in the pseudorapidity range $2.0 < \eta < 4.5$, $\pt(\mu+j) > 20\gev$, and $\Delta R(\mu, j) > 0.5$.  For $Z+$jet events both muons must fulfill the muon requirements and $60 < M(\mu\mu) < 120\gev$; the $Z+$jet fiducial region does not require $\pt(\mu+j) > 20\gev$.
}
    \begin{tabular}{>$l<$>$r<$>$r<$>$r<$>$r<$}
      \toprule
      & \multicolumn{2}{c}{Results} & \multicolumn{2}{c}{SM prediction} \\
      & \multicolumn{1}{c}{$7\tev$} & \multicolumn{1}{c}{$8\tev$} & \multicolumn{1}{c}{$7\tev$} & \multicolumn{1}{c}{$8\tev$} \\
      \midrule
      \frac{\sigma(Wb)}{\sigma(Wj)} \times 10^2 & 0.66\pm0.13\pm0.13 & 0.78\pm0.08\pm0.16 
& 0.74_{-0.13}^{+0.17} & 0.77_{-0.13}^{+0.18} \\
      \frac{\sigma(Wc)}{\sigma(Wj)} \times 10^2& 5.80\pm0.44\pm0.75 & 5.62\pm0.28\pm0.73 
& 5.02_{-0.69}^{+0.80} & 5.31_{-0.52}^{+0.87} \\
      \midrule
      \mathcal{A}(Wb) & 0.51\pm0.20\pm0.09 & 0.27\pm0.13\pm0.09 
& 0.27_{-0.03}^{+0.03} & 0.28_{-0.03}^{+0.03} \\
      \mathcal{A}(Wc) & -0.09\pm0.08\pm0.04 & -0.01\pm0.05\pm0.04 
& -0.15_{-0.04}^{+0.02} & -0.14_{-0.03}^{+0.02} \\
      \midrule
      \frac{\sigma(W^+j)}{\sigma(Zj)} & 10.49 \pm 0.28\pm 0.53 & 9.44 \pm 0.19\pm 0.47 
& 9.90_{-0.24}^{+0.28} & 9.48_{-0.33}^{+0.16} \\
      \frac{\sigma(W^-j)}{\sigma(Zj)} & 6.61 \pm 0.19\pm 0.33 & 6.02 \pm 0.13\pm 0.30 
& 5.79_{-0.18}^{+0.21} & 5.52_{-0.25}^{+0.13} \\
      \bottomrule
    \end{tabular}
  \end{center}
\end{table}

\section{Results}\label{sec:results}

The results for $\sqrt{s} = 7$ and 8\tev are summarized in Table~\ref{tab:summary}.   Each result is compared to SM predictions calculated at NLO using MCFM~\cite{Campbell:2000bg} and the CT10 PDF set~\cite{Lai:2010vv} as described in Sect.~\ref{sec:data}.  
Production of $W+$jet events in the forward region requires a large imbalance in $x$ of the initial partons.  
In the four-flavor scheme at leading order, \wb production proceeds via $q\bar{q}\to Wg(b\bar{b})$, where the charge of the $W$ boson has the same sign as that of the initial parton with larger $x$.  
Therefore, $\mathcal{A}(Wb) \approx +1/3$ is predicted due to the valence quark content of the proton. 
The dominant mechanism for \wc production is $gs \to Wc$, which is charge symmetric assuming symmetric $s$ and $\bar{s}$ quark PDFs.  However, the Cabibbo-suppressed contribution from $gd\to Wc$ leads to a prediction of a small negative value for $\mathcal{A}(Wc)$.

The $\sigma(Wb)/\sigma(Wj)$ ratio in conjunction with the \wb charge asymmetry is consistent with MCFM calculations performed in the four-flavor scheme, where  \wb production is primarily from gluon splitting.   This scheme assumes no intrinsic $b$ quark content in the proton.  
The data do not support a large contribution from intrinsic $b$ quark content in the proton but the precision is not sufficient to rule out such a contribution at $\mathcal{O}(10\%)$. 
The ratio $[\sigma(Wb)+\sigma({\rm top})]/\sigma(Wj)$ %, {\em i.e.} the ratio for the \wb final state without top quark subtraction, 
is measured to be $1.17\pm0.13\stat\pm0.18\syst$\% at $\sqrt{s}=7$\tev and $1.29\pm0.08\stat\pm0.19\syst$\% at $\sqrt{s}=8$\tev, which agree with the  NLO SM predictions of $1.23\pm0.24\%$ and $1.38\pm0.26\%$, respectively.  

The $\sigma(Wc)/\sigma(Wj)$ ratio is much larger than $\sigma(Wb)/\sigma(Wj)$, which is consistent with $Wc$ production from intrinsic $s$ quark content of the proton.  
The measured charge asymmetry for \wc is about $2\sigma$ smaller than the predicted value obtained with CT10, which assumes symmetric $s$ and $\bar{s}$ quark PDFs.  
This could suggest a larger than expected contribution from scattering off of strange quarks or a charge asymmetry between $s$ and $\bar{s}$ quarks in the proton.  
The ratio $\sigma(W^+j)/\sigma(Zj)$ is consistent within $1\sigma$ with NLO predictions, while the observed $\sigma(W^-j)/\sigma(Zj)$ ratio is higher than the predicted value by about $1.5\sigma$.

\section*{Acknowledgments}

\noindent We express our gratitude to our colleagues in the CERN
accelerator departments for the excellent performance of the LHC. We
thank the technical and administrative staff at the LHCb
institutes. We acknowledge support from CERN and from the national
agencies: CAPES, CNPq, FAPERJ and FINEP (Brazil); NSFC (China);
CNRS/IN2P3 (France); BMBF, DFG, HGF and MPG (Germany); INFN (Italy); 
FOM and NWO (The Netherlands); MNiSW and NCN (Poland); MEN/IFA (Romania); 
MinES and FANO (Russia); MinECo (Spain); SNSF and SER (Switzerland); 
NASU (Ukraine); STFC (United Kingdom); NSF (USA).
The Tier1 computing centres are supported by IN2P3 (France), KIT and BMBF 
(Germany), INFN (Italy), NWO and SURF (The Netherlands), PIC (Spain), GridPP 
(United Kingdom).
We are indebted to the communities behind the multiple open 
source software packages on which we depend. We are also thankful for the 
computing resources and the access to software R\&D tools provided by Yandex LLC (Russia).
Individual groups or members have received support from 
EPLANET, Marie Sk\l{}odowska-Curie Actions and ERC (European Union), 
Conseil g\'{e}n\'{e}ral de Haute-Savoie, Labex ENIGMASS and OCEVU, 
R\'{e}gion Auvergne (France), RFBR (Russia), XuntaGal and GENCAT (Spain), Royal Society and Royal
Commission for the Exhibition of 1851 (United Kingdom).

\addcontentsline{toc}{section}{References}
\setboolean{inbibliography}{true}
\bibliographystyle{LHCb}
\bibliography{wbc_paper.bbl}

\newpage

% Author List ----------------------------
%\input{LHCb_HD_authorlist_2015-03-31.tex}
%%%%%%%%%%%%%%%%%%%%%%%%%%%%%%%%%%%%%%%%%%
\centerline{\large\bf LHCb collaboration}
\begin{flushleft}
\small
R.~Aaij$^{38}$, 
B.~Adeva$^{37}$, 
M.~Adinolfi$^{46}$, 
A.~Affolder$^{52}$, 
Z.~Ajaltouni$^{5}$, 
S.~Akar$^{6}$, 
J.~Albrecht$^{9}$, 
F.~Alessio$^{38}$, 
M.~Alexander$^{51}$, 
S.~Ali$^{41}$
% ...

{\footnotesize \it
% ...
$ ^{5}$Clermont Universit\'{e}, Universit\'{e} Blaise Pascal, CNRS/IN2P3, LPC, Clermont-Ferrand, France\\
$ ^{6}$CPPM, Aix-Marseille Universit\'{e}, CNRS/IN2P3, Marseille, France\\
% ...
$ ^{9}$Fakult\"{a}t Physik, Technische Universit\"{a}t Dortmund, Dortmund, Germany\\
% ...
$ ^{37}$Universidad de Santiago de Compostela, Santiago de Compostela, Spain\\
$ ^{38}$European Organization for Nuclear Research (CERN), Geneva, Switzerland\\
% ...
$ ^{41}$Nikhef National Institute for Subatomic Physics, Amsterdam, The Netherlands\\
% ...
$ ^{46}$H.H. Wills Physics Laboratory, University of Bristol, Bristol, United Kingdom\\
% ...
$ ^{51}$School of Physics and Astronomy, University of Glasgow, Glasgow, United Kingdom\\
$ ^{52}$Oliver Lodge Laboratory, University of Liverpool, Liverpool, United Kingdom\\
% ...
}
\end{flushleft}
%%%%%%%%%%%%%%%%%%%%%%%%%%%%%%%%%%%%%%%%%%

\end{document}